\documentclass[preprint,aps,showpacs,showkeys,superscriptaddress]{revtex4}%
\usepackage[latin9]{inputenc}
\usepackage{amsmath}
\usepackage{amssymb}
\usepackage{graphicx}
\usepackage{amsfonts}
\usepackage{subfig}%
\makeatletter
\@ifundefined{textcolor}{}
{
\definecolor{BLACK}{gray}{0}
 \definecolor{WHITE}{gray}{1}
 \definecolor{RED}{rgb}{1,0,0}
 \definecolor{GREEN}{rgb}{0,1,0}
 \definecolor{BLUE}{rgb}{0,0,1}
 \definecolor{CYAN}{cmyk}{1,0,0,0}
 \definecolor{MAGENTA}{cmyk}{0,1,0,0}
 \definecolor{YELLOW}{cmyk}{0,0,1,0}
}
\@ifundefined{textcolor}{}{
\definecolor{BLACK}{gray}{0}
 \definecolor{WHITE}{gray}{1}
 \definecolor{RED}{rgb}{1,0,0}
 \definecolor{GREEN}{rgb}{0,1,0}
 \definecolor{BLUE}{rgb}{0,0,1}
 \definecolor{CYAN}{cmyk}{1,0,0,0}
 \definecolor{MAGENTA}{cmyk}{0,1,0,0}
 \definecolor{YELLOW}{cmyk}{0,0,1,0}
}
\@ifundefined{textcolor}{}{
\definecolor{BLACK}{gray}{0}
 \definecolor{WHITE}{gray}{1}
 \definecolor{RED}{rgb}{1,0,0}
 \definecolor{GREEN}{rgb}{0,1,0}
 \definecolor{BLUE}{rgb}{0,0,1}
 \definecolor{CYAN}{cmyk}{1,0,0,0}
 \definecolor{MAGENTA}{cmyk}{0,1,0,0}
 \definecolor{YELLOW}{cmyk}{0,0,1,0}
}
\@ifundefined{textcolor}{}{
\definecolor{BLACK}{gray}{0}
 \definecolor{WHITE}{gray}{1}
 \definecolor{RED}{rgb}{1,0,0}
 \definecolor{GREEN}{rgb}{0,1,0}
 \definecolor{BLUE}{rgb}{0,0,1}
 \definecolor{CYAN}{cmyk}{1,0,0,0}
 \definecolor{MAGENTA}{cmyk}{0,1,0,0}
 \definecolor{YELLOW}{cmyk}{0,0,1,0}
}
\@ifundefined{textcolor}{}{
\definecolor{BLACK}{gray}{0}
 \definecolor{WHITE}{gray}{1}
 \definecolor{RED}{rgb}{1,0,0}
 \definecolor{GREEN}{rgb}{0,1,0}
 \definecolor{BLUE}{rgb}{0,0,1}
 \definecolor{CYAN}{cmyk}{1,0,0,0}
 \definecolor{MAGENTA}{cmyk}{0,1,0,0}
 \definecolor{YELLOW}{cmyk}{0,0,1,0}
}
\@ifundefined{textcolor}{}{
\definecolor{BLACK}{gray}{0}
 \definecolor{WHITE}{gray}{1}
 \definecolor{RED}{rgb}{1,0,0}
 \definecolor{GREEN}{rgb}{0,1,0}
 \definecolor{BLUE}{rgb}{0,0,1}
 \definecolor{CYAN}{cmyk}{1,0,0,0}
 \definecolor{MAGENTA}{cmyk}{0,1,0,0}
 \definecolor{YELLOW}{cmyk}{0,0,1,0}
}
\@ifundefined{textcolor}{}{
\definecolor{BLACK}{gray}{0}
 \definecolor{WHITE}{gray}{1}
 \definecolor{RED}{rgb}{1,0,0}
 \definecolor{GREEN}{rgb}{0,1,0}
 \definecolor{BLUE}{rgb}{0,0,1}
 \definecolor{CYAN}{cmyk}{1,0,0,0}
 \definecolor{MAGENTA}{cmyk}{0,1,0,0}
 \definecolor{YELLOW}{cmyk}{0,0,1,0}
}
\DeclareTextSymbolDefault{\textquotedbl}{T1}
\@ifundefined{textcolor}{}{
\definecolor{BLACK}{gray}{0}
 \definecolor{WHITE}{gray}{1}
 \definecolor{RED}{rgb}{1,0,0}
 \definecolor{GREEN}{rgb}{0,1,0}
 \definecolor{BLUE}{rgb}{0,0,1}
 \definecolor{CYAN}{cmyk}{1,0,0,0}
 \definecolor{MAGENTA}{cmyk}{0,1,0,0}
 \definecolor{YELLOW}{cmyk}{0,0,1,0}
}
\DeclareTextSymbolDefault{\textquotedbl}{T1}
\@ifundefined{textcolor}{}{
\definecolor{BLACK}{gray}{0}
 \definecolor{WHITE}{gray}{1}
 \definecolor{RED}{rgb}{1,0,0}
 \definecolor{GREEN}{rgb}{0,1,0}
 \definecolor{BLUE}{rgb}{0,0,1}
 \definecolor{CYAN}{cmyk}{1,0,0,0}
 \definecolor{MAGENTA}{cmyk}{0,1,0,0}
 \definecolor{YELLOW}{cmyk}{0,0,1,0}
}
\DeclareTextSymbolDefault{\textquotedbl}{T1}
\@ifundefined{textcolor}{}{
\definecolor{BLACK}{gray}{0}
 \definecolor{WHITE}{gray}{1}
 \definecolor{RED}{rgb}{1,0,0}
 \definecolor{GREEN}{rgb}{0,1,0}
 \definecolor{BLUE}{rgb}{0,0,1}
 \definecolor{CYAN}{cmyk}{1,0,0,0}
 \definecolor{MAGENTA}{cmyk}{0,1,0,0}
 \definecolor{YELLOW}{cmyk}{0,0,1,0}
}
\DeclareTextSymbolDefault{\textquotedbl}{T1}
\@ifundefined{textcolor}{}{
\definecolor{BLACK}{gray}{0}
 \definecolor{WHITE}{gray}{1}
 \definecolor{RED}{rgb}{1,0,0}
 \definecolor{GREEN}{rgb}{0,1,0}
 \definecolor{BLUE}{rgb}{0,0,1}
 \definecolor{CYAN}{cmyk}{1,0,0,0}
 \definecolor{MAGENTA}{cmyk}{0,1,0,0}
 \definecolor{YELLOW}{cmyk}{0,0,1,0}
}
\DeclareTextSymbolDefault{\textquotedbl}{T1}
\@ifundefined{textcolor}{}{
\definecolor{BLACK}{gray}{0}
 \definecolor{WHITE}{gray}{1}
 \definecolor{RED}{rgb}{1,0,0}
 \definecolor{GREEN}{rgb}{0,1,0}
 \definecolor{BLUE}{rgb}{0,0,1}
 \definecolor{CYAN}{cmyk}{1,0,0,0}
 \definecolor{MAGENTA}{cmyk}{0,1,0,0}
 \definecolor{YELLOW}{cmyk}{0,0,1,0}
}
\DeclareTextSymbolDefault{\textquotedbl}{T1}
\@ifundefined{textcolor}{}{
\definecolor{BLACK}{gray}{0}
 \definecolor{WHITE}{gray}{1}
 \definecolor{RED}{rgb}{1,0,0}
 \definecolor{GREEN}{rgb}{0,1,0}
 \definecolor{BLUE}{rgb}{0,0,1}
 \definecolor{CYAN}{cmyk}{1,0,0,0}
 \definecolor{MAGENTA}{cmyk}{0,1,0,0}
 \definecolor{YELLOW}{cmyk}{0,0,1,0}
}

\@ifundefined{showcaptionsetup}{}{ \PassOptionsToPackage{caption=false}{subfig}}
\@ifundefined{showcaptionsetup}{}{ \PassOptionsToPackage{caption=false}{subfig}}
\@ifundefined{showcaptionsetup}{}{ \PassOptionsToPackage{caption=false}{subfig}}
\@ifundefined{showcaptionsetup}{}{ \PassOptionsToPackage{caption=false}{subfig}}
\@ifundefined{showcaptionsetup}{}{ \PassOptionsToPackage{caption=false}{subfig}}
\@ifundefined{showcaptionsetup}{}{ \PassOptionsToPackage{caption=false}{subfig}}
\@ifundefined{showcaptionsetup}{}{ \PassOptionsToPackage{caption=false}{subfig}}
\@ifundefined{showcaptionsetup}{}{ \PassOptionsToPackage{caption=false}{subfig}}
\@ifundefined{showcaptionsetup}{}{ \PassOptionsToPackage{caption=false}{subfig}}
\@ifundefined{showcaptionsetup}{}{ \PassOptionsToPackage{caption=false}{subfig}}
\@ifundefined{showcaptionsetup}{}{ \PassOptionsToPackage{caption=false}{subfig}}
\@ifundefined{showcaptionsetup}{}{ \PassOptionsToPackage{caption=false}{subfig}}
\@ifundefined{showcaptionsetup}{}{ \PassOptionsToPackage{caption=false}{subfig}}
\makeatother
\begin{document}
\title{{\Large {}{}{}{}{}{}{}{}{}{}{}{}{}Generalized para-Bose states}}
\author{A. S. Pereira}
\email{alfaspereira@gmail.com}
\affiliation{Instituto Federal do Par\'{a}, 68600-000 Bragan\c{c}a, Par\'{a}, Brazil }
\affiliation{Departamento de F\'{\i}sica, Universidade Federal de Campina Grande, Caixa
Postal 10071, 58429-900 Campina Grande, Para\'{\i}ba, Brazil}
\author{A. S. Lemos}
\email{adiellemos@gmail.com}
\affiliation{Departamento de F\'{\i}sica, Universidade Federal de Campina Grande, Caixa
Postal 10071, 58429-900 Campina Grande, Para\'{\i}ba, Brazil}
\author{F. A. Brito}
\email{fabrito@df.ufcg.edu.br}
\affiliation{Departamento de F\'{\i}sica, Universidade Federal de Campina Grande, Caixa
Postal 10071, 58429-900 Campina Grande, Para\'{\i}ba, Brazil}
\affiliation{Departamento de F\'{\i}sica, Universidade Federal da Para\'{\i}ba, Caixa
Postal 5008, 58051-970 Jo\~{a}o Pessoa, Para\'{\i}ba, Brazil}


\begin{abstract}
In this paper, we construct integrals of motion in a para-Bose formulation for
a general time-dependent quadratic Hamiltonian, which, in its turn, commutes
with the reflection operator. In this context, we obtain generalizations for
the squeezed vacuum states (SVS) and coherent states (CS) in terms of the
Wigner parameter. Furthermore, we show that there is a completeness relation
for the generalized SVS owing to the Wigner parameter. In the study of the
probability transition, we found that the displacement parameter acts as a
transition parameter by allowing access to odd states, while the Wigner
parameter controls the dispersion of the distribution. We show that the Wigner
parameter is quantized by imposing that the vacuum state has even parity. We
apply the general results to the case of the time-independent para-Bose
oscillator and find that the mean values of the coordinate and momentum have
an oscillatory behavior similarly to the simple harmonic oscillator, while the
standard deviation presents corrections in terms of the squeeze, displacement,
and Wigner parameters.

\end{abstract}
\maketitle

\section{Introduction}

In 1926 Schr\"{o}dinger obtained quantum states to harmonic oscillator, which
allowed for a semiclassical description \citep{SCH1926}. In the 1960s, with
the advent of the laser, these states were rediscovered and widespread in the
works of Glauber-Sudarshan-Klauber \citep{GLA1963,SUD1963,KLA1963} by showing
a semiclassical description of electromagnetic radiation, and where the term
``coherent states" (CS){} was presented for the first time. These states for
the harmonic oscillator can be obtained in three equivalent ways: I - as
eigenstates of the annihilation operator; II - by the action of a unitary
displacement operator acting in the vacuum state; III - as minimum uncertainty
states. The minimization of the uncertainty relation coincides with the value
calculated in the vacuum state of the harmonic oscillator, which presents the
same value for the standard deviation in both position and momentum. The
states satisfying these properties are called canonical CS. In addition, these
states have properties of continuity and form an overcomplete basis, which
guarantees a prominent role in modern quantum mechanics, with various
applications ranging from quantum optics \citep{SCU1997,KLA1968}, quantum
computing \citep{NIE2000}, and mathematical physics \citep{KLA1985}.

In its turn, squeezed states (SS) form a class of nonclassical states, which
minimize the uncertainty relation \cite{Bac2004,SCH2017,SAN2021}. At the same
time, these states offer the possibility of reducing the standard deviation of
a physical quantity to a value less than that calculated in the vacuum state
at the expense of increasing the standard deviation of other physical quantity
\cite{Sto1970,Sto1971}. Thus, these states have great potential for
application, for example, in quantum information \cite{Slu1990} and for the
detectors of gravitational waves \cite{Cav1981,Ni1987,Chu2014}. Usually, these
states are constructed by the action of the unitary squeeze operator on an
arbitrary state. In particular, the acting of the squeeze operator on the
vacuum state leads to the squeezed vacuum states (SVS) \cite{Ger1983}.

Generalizing quantum states provides a deeper understanding of corresponding
systems since they sometimes bring additional degrees of freedom. In this
sense, there is a wide range of publications in the literature that seeks to
generalize CS to be able to describe systems in addition to the harmonic
oscillator, such as the time-dependent quadratic systems
\citep{MAL1979,DOD2003}, systems with a given Lie group
\citep{PER1986,GIL1972}, some non-trivial generalizations \citep{ALI2000}, and
squeezed coherent states (SCS) \citep{YUE1976,SAT1985}. On the other hand, CS
can be generalized by considering deformations in the canonical commutation
relation and thus allow us to study a wide range of relevant problems. For
instance, the implications of the gravitational effects in quantum mechanics
\citep{DEY2012,PED2013,PAT2021}, q- and f-deformed oscillators study which is
applied, mainly, to quantum optics \citep{MAN1997,SAN2015,ERE2006}, and to
study systems with singularity as the Calogero-like model via
Wigner-Heisenberg algebra (WHA) (or R-deformed Heisenberg algebra)
\cite{Aga1995,BRI1993}. In particular, we are interested in generalizing CS
and SVS via WHA.

The WHA originated from the quantization proposed by Wigner \citep{WIG1950},
which generalizes the canonical commutation relation. In particular, this
algebra is generated by the creation-annihilation operators and the reflection
operator (parity operator), which satisfies commutation and anti-commutation
relations \citep{MIK1997}. In this formulation, it is possible to obtain a
self-adjoint momentum operator on the semi-axis in terms of a parameter that
deforms the canonical commutation relation \cite{MUK1980,LOH2004}.
Furthermore, the WHA obeys the trilinear commutation and anti-commutation
relations, which characterize the para-Bose operators \cite{SHA1981}. On its
turn, this algebra leads to the parastatistical description of physical
systems in terms of a deformation parameter, which corresponds to
generalization of the Bose-Einstein and Fermi-Dirac statistics \citep{HAR1968,GRE1953,OHN1982}.

According to the order of the deformation parameter, also known as the order
of statistics, it is possible to describe para-Bose or para-Fermi particles,
see also \citep{STO2010}. Although parastatistics cannot be applied to the
particles described by the standard model \citep{GRE1965}, their distinctive
approach has allowed presenting promising proposals, for instance, in the
conjecture of candidate particles to explain dark matter \citep{KIT2018}, in
paraquark models description \citep{BRA1973}, in the study of thermodynamics
properties of para-Bose systems \citep{HAM1992}, and in optical physics by
simulation of para-Fermi oscillators \citep{WAL2020}. By considering the
para-Bose formulation, we will construct integrals of motion, i.e., operators
that commute with the Schr\"{o}dinger's operator \citep{PER2021} in the form
of a Bogoliubov transformation \citep{MEY1980}.

In the para-Bose formulation \citep{JOR1963}, the CS were obtained as
eigenstates of the annihilation operator that satisfy the characteristic
commutation relation of the algebra \citep{SHA1978} -- see
\citep{SHA1981,BIS1980,SAX1986}. Recently, some works have paid attention to
this topic, for instance, the nonlinear CS \citep{MOJ2018a}, the construction
of new types of para-Bose states \cite{MOJ2018b,Ald2017,Bag1997}, and the
study of ``Schr\"{o}dinger cat states''\ \citep{DEH2015}. In this context, we
will obtain the time-dependent generalized CS and SVS via integrals of motion.
These states can be applied to describe Calogero-like models \cite{CAL1971}.
In its turn, the Calogero model and its generalizations are able to describe
several physical phenomena \cite{SIM1994}, such as the Hall effect
\cite{KAW1993,AZU1994}, anyons \cite{OUV2001}, and fluctuations in mesoscopic
systems \cite{CAS1995}, for instance. Furthermore, our construction has
potential application to describe the recent proposed experimental realization
of para-particles \cite{ALD2021,ALD2017}.

This paper is structured as follows. In Sec. \ref{II}, the integrals of motion
method in the context of WHA will be formulated. In Sec. \ref{III}, the
time-dependent generalized SVS in terms of the time-independent para-Bose
number states will be constructed. In turn, the completeness relation and
probability transition plots will be obtained. In Sec. \ref{IV}, a
generalization of the time-dependent CS in terms of the Wigner, squeeze and
displacement parameters will be constructed. Then, the probability transition
graph will be shown, and the mean values and the uncertainty relations will be
calculated. In Sec. \ref{V}, the generalized CS in the coordinate
representation will be considered. We apply the general result to the
particular case of the para-Bose oscillator in Sec. \ref{VI}. Then, the
concluding remarks are presented in Sec. \ref{VII}.

\section{Integrals of motion via WHA\label{II}}

The motion integral method consists of building a time-dependent operator,
which commutes with the Schr\"{o}dinger's operator. In turn, the eigenstates
of this operator are obtained and imposed to satisfy the Schr\"{o}dinger's
equation. This technique introduced by Lewis and Reisenfield to study the
time-dependent harmonic oscillator \cite{Lew1969} proved to be useful in
several other problems, e.g., in the study of time-dependent quadratic
Hamiltonians in one dimension \cite{DOD1975,Bag2015,And1999} and for
multidimensional systems \cite{Dod1989}, Dirac equation \cite{Man1997}, and
supersymmetric quantum mechanics \cite{Mos2001}. Here, our contribution will
be to reformulate the integrals of motion method according to the WHA. This
algebraic formulation allows studying time-dependent systems with a
singularity at the origin characterized by centrifugal potentials.

The WHA is composed by the annihilation $\hat{a}$, creation $\hat{a}^{\dagger
}$ and reflection $\hat{R}$ operators, satisfying commutation $\left(  \left[
\hat{b},\hat{c}\right]  =\hat{b}\hat{c}-\hat{c}\hat{b}\right)  $ and
anti-commutation $\left(  \left\{  \hat{b},\hat{c}\right\}  =\hat{b}\hat
{c}+\hat{c}\hat{b}\right)  $ relations,
\begin{equation}
\left[  \hat{a},\hat{a}^{\dagger}\right]  =1+\nu\hat{R},\text{ \ }\left\{
\hat{R},\hat{a}\right\}  =0=\left\{  \hat{R},\hat{a}^{\dagger}\right\}
,\text{ \ }\hat{R}^{2}=1, \label{1}%
\end{equation}
where $\nu$ is the Wigner parameter (deformation parameter) related to the
fundamental energy level $\varepsilon$ of the para-Bose oscillator given by
\begin{equation}
\nu=2\varepsilon-1. \label{2}%
\end{equation}
Notice that by setting $\nu=0\Rightarrow\varepsilon=1/2$ we recover the
canonical commutation relation.

The operators $\hat{a}$ and $\hat{a}^{\dagger}$ also obey the trilinear
commutation and anti-commutation relation
\begin{equation}
\left[  \left\{  \hat{a},\hat{a}^{\dagger}\right\}  ,\hat{a}\right]
=-2\hat{a},\text{ \ }\left[  \left\{  \hat{a},\hat{a}^{\dagger}\right\}
,\hat{a}^{\dagger}\right]  =2\hat{a}^{\dagger}, \label{2-1}%
\end{equation}
which characterize the para-Bose operators \cite{MIK1997,SHA1981}.

For the sake of simplicity, we will consider a Hamiltonian which commutes with
the reflection operator, and in that way, the eigenstates of $\hat{H}$ can be
even or odd. In this sense, the most general form for a one-dimensional
time-dependent quadratic Hamiltonian is given by
\begin{equation}
\hat{H}=\frac{1}{2}\hbar\left(  \alpha^{\ast}\hat{a}^{2}+\alpha\hat
{a}^{\dagger2}\right)  +\frac{1}{2}\hbar\beta\left(  \hat{a}^{\dagger}\hat
{a}+\hat{a}\hat{a}^{\dagger}\right)  +\hbar\delta, \label{3}%
\end{equation}
where $\alpha=\alpha\left(  t\right)  $, $\beta=\beta\left(  t\right)  $ and
$\delta=\delta\left(  t\right)  $ are time-dependent functions and the signs
$\dagger$ and $\ast$ denote Hermitian and complex conjugation, respectively.
From the hermiticity condition $\hat{H}=\hat{H}^{\dagger}$, we have that
$\beta$ and $\delta$ must be real functions. By assuming the condition
$\beta>\left\vert \alpha\right\vert $, we get that the Hamiltonian (\ref{3})
is positive definite, i.e., it can be written in the form of an
oscillator-type Hamiltonian.

In its turn, the quantum states $\left\vert \Psi\right\rangle $ which describe
the time evolution of the system should satisfy the Schr\"{o}dinger's
equation
\begin{align}
&  \hat{\Lambda}\left\vert \Psi\right\rangle =0,\text{ \ }\nonumber\\
&  \hat{\Lambda}=\frac{i}{\hbar}\hat{H}+\partial_{t},\text{ \ }\partial
_{t}=\frac{\partial}{\partial t}, \label{4}%
\end{align}
where $\hat{\Lambda}$ is the Schr\"{o}dinger's operator.

Let us consider a time-dependent operator $\hat{A}=\hat{A}\left(  t\right)  $,
as a linear combination of the $\hat{a}$ and $\hat{a}^{\dagger}$ operators, in
the form
\begin{equation}
\hat{A}=f\hat{a}+g\hat{a}^{\dagger}+\varphi, \label{5}%
\end{equation}
where $f=f\left(  t\right)  $, $g=g\left(  t\right)  $ and $\varphi
=\varphi\left(  t\right)  $ are time-dependent complex functions. The Eq.
(\ref{5}) is the well-known Bogoliubov's transformation; those coefficients
have been studied in Ref. \citep{MEY1980}.

In order to $\hat{A}$ to be an integral of motion, it has to commute with the
Schr\"{o}dinger's operator (\ref{4}), that is
\begin{equation}
\overset{\cdot}{\hat{A}}=\left[  \hat{\Lambda},\hat{A}\right]  =0,\text{
\ }\overset{\cdot}{\hat{A}}=\frac{d\hat{A}}{dt}. \label{6}%
\end{equation}
Substituting the Eqs. (\ref{3}), (\ref{4}) and (\ref{5}) into (\ref{6}), we
obtain the following equations for the functions $f$, $g$ and $\varphi$:
\begin{equation}
\dot{f}=i\left(  \beta f-\alpha^{\ast}g\right)  ,\text{ \ }\dot{g}=i\left(
f\alpha-\beta g\right)  ,\text{ \ }\dot{\varphi}=0, \label{7}%
\end{equation}
where $\varphi\left(  t\right)  =\varphi_{0}$ is a constant function for any
instant of time.

The commutator between $\hat{A}$ and $\hat{A}^{\dagger}$ reads
\begin{equation}
\left[  \hat{A},\hat{A}^{\dagger}\right]  =\mu\left(  1+\nu\hat{R}\right)
,\text{\ }\mu=\left\vert f\right\vert ^{2}-\left\vert g\right\vert
^{2}=\left\vert f_{0}\right\vert ^{2}-\left\vert g_{0}\right\vert
^{2},\text{\ }\text{for arbitrary $t$}, \label{7a}%
\end{equation}
where $f_{0}=f\left(  0\right)  $ and $g_{0}=g\left(  0\right)  $ are initial conditions.

It follows from (\ref{5}) and (\ref{7a}), that
\begin{equation}
\hat{a}=\frac{f^{\ast}\hat{A}-g\hat{A}^{\dagger}+u}{\mu},\text{\ }%
u=g\varphi_{0}^{\ast}-f^{\ast}\varphi_{0}. \label{8}%
\end{equation}
Notice that $\left\vert f\right\vert \neq\left\vert g\right\vert $ for all
time ensures that $\hat{a}$ is well defined for all $t$. It is worth
highlighting that the $\mu$-parameter will be useful to study special cases in
which the Hamiltonian presents linear terms. In addition, the $\mu$-parameter
will be useful to replace the functions $f$ and $g$ with the squeeze and
displacement parameters in the constructed states.

\section{Time-dependent generalized SVS\label{III}}

The SVS are pure states known to be one of the most useful nonclassical
states, see for instance, \cite{SCH2017}. In addition, these states make it
possible to obtain the standard deviation of a physical quantity less than its
value in the vacuum state. In this section, we will obtain the time-dependent
generalized SVS via para-Bose formulation. Thus, assuming the condition
\begin{equation}
\varphi_{0}=0\Longrightarrow u=0, \label{8a}%
\end{equation}
one can obtain the generalized SVS following the nonunitary approach, as
described in Ref. \cite{PER2021}. In what follows we will apply this condition.

\subsection{Para-Bose number states}

In this subsection, we will recall some properties of the para-Bose number
states, since the generalized SVS can be expanded in these states. It is
well-known that the para-Bose number states $\left\vert n,\varepsilon
\right\rangle $ form a complete set and are orthonormal
\citep{SHA1978,SHA1981}, i.e.,
\begin{equation}%
{\displaystyle\sum\limits_{n=0}^{\infty}}
\left\vert n,\varepsilon\right\rangle \left\langle \varepsilon,n\right\vert
=1,\text{ \ }\left\langle \varepsilon,n|m,\varepsilon\right\rangle
=\delta_{n,m}, \label{8b}%
\end{equation}
where $\delta_{n,m}$ is the Kronecker delta.

Taking into account that the vacuum state has even parity $\left(  \hat
{R}\left\vert 0,\varepsilon\right\rangle =\left\vert 0,\varepsilon
\right\rangle \right)  $, the action of the generators of the WHA in
$\left\vert n,\varepsilon\right\rangle $, reads
\begin{align}
&  \hat{a}\left\vert 2n,\varepsilon\right\rangle =\sqrt{2n}\left\vert
2n-1,\varepsilon\right\rangle ,\text{ \ }\hat{a}^{\dagger}\left\vert
2n,\varepsilon\right\rangle =\sqrt{2\left(  n+\varepsilon\right)  }\left\vert
2n+1,\varepsilon\right\rangle ,\nonumber\\
&  \hat{a}\left\vert 2n+1,\varepsilon\right\rangle =\sqrt{2\left(
n+\varepsilon\right)  }\left\vert 2n,\varepsilon\right\rangle ,\text{ \ }%
\hat{a}^{\dagger}\left\vert 2n+1,\varepsilon\right\rangle =\sqrt{2\left(
n+1\right)  }\left\vert 2n+2,\varepsilon\right\rangle ,\nonumber\\
&  \hat{n}\left\vert n,\varepsilon\right\rangle =\left(  \frac{1}{2}\left\{
\hat{a},\hat{a}^{\dagger}\right\}  -\varepsilon\right)  \left\vert
n,\varepsilon\right\rangle =n\left\vert n,\varepsilon\right\rangle ,\text{
\ }\hat{R}\left\vert n,\varepsilon\right\rangle =\left(  -1\right)
^{n}\left\vert n,\varepsilon\right\rangle ,\text{ \ }n=0,1,2,3,\ldots.
\label{9}%
\end{align}
From a recurrence relation, one can write the number states $\left\vert
n,\varepsilon\right\rangle $ in terms of the vacuum state $\left\vert
0,\varepsilon\right\rangle $, in the form
\begin{align}
&  \left\vert 2n,\varepsilon\right\rangle =\sqrt{\frac{\Gamma\left(
\varepsilon\right)  }{2^{2n}n!\Gamma\left(  n+\varepsilon\right)  }}\left(
\hat{a}^{\dagger}\right)  ^{2n}\left\vert 0,\varepsilon\right\rangle
,\nonumber\\
&  \left\vert 2n+1,\varepsilon\right\rangle =\sqrt{\frac{\Gamma\left(
\varepsilon\right)  }{2^{2n+1}n!\Gamma\left(  n+\varepsilon+1\right)  }%
}\left(  \hat{a}^{\dagger}\right)  ^{2n+1}\left\vert 0,\varepsilon
\right\rangle . \label{10}%
\end{align}

\subsection{Time-dependent generalized SVS via para-Bose number states}

In what follows, we aim to apply the nonunitary approach, as in the recent
publication \citep{PER2021}, to construct the generalized SVS. From condition
({\ref{8a}}), one can write the nonunitary operator $\hat{S}$, in the form
\begin{equation}
\hat{S}=\exp\left(  \frac{1}{2}\zeta\hat{a}^{\dagger2}\right)  ,\text{
\ }\zeta=\frac{g}{f},\text{ \ }\dot{\zeta}=i\alpha^{\ast}\zeta^{2}%
-2i\beta\zeta+i\alpha, \label{11}%
\end{equation}
such that the commutators from Baker-Campbell-Hausdorff relation of the second
order onwards become null. Here, $\zeta=\zeta\left(  t\right)  $ represents
the squeeze parameter for the generalized SVS. The form of $\hat{S}$ allows us
to introduce the most convenient squeeze parameter, as well as relate the
motion integral $\hat{A}$ directly with the annihilation operator $\hat{a}$.
Furthermore, this approach guarantees that there is a direct relationship
between the eigenstates of the motion integral with the para-Bose number
states, as we will see next.

Applying the Baker--Campbell--Hausdorff theorem, we can express the operator
$\hat{a}$ in terms of the integrals of motion $\hat{A}$, as follows
\begin{equation}
\hat{a}=\frac{1}{f}\hat{S}\hat{A}\hat{S}^{-1}. \label{14}%
\end{equation}

The application from (\ref{14}) on the vacuum state $\left\vert 0,\varepsilon
\right\rangle $, which satisfies the annihilation condition $\hat{a}\left\vert
0,\varepsilon\right\rangle =0$, yields:
\begin{equation}
\hat{A}\left\vert \zeta\right\rangle =0,\text{ } \label{15}%
\end{equation}
whose general solution is given by
\begin{equation}
\text{\ }\left\vert \zeta\right\rangle =\Phi\exp\left(  -\frac{1}{2}\zeta
\hat{a}^{\dagger2}\right)  \left\vert 0,\varepsilon\right\rangle , \label{16}%
\end{equation}
where $\Phi=\Phi\left(  t\right)  $ is an arbitrary function, which will be
determined such that the states $\left\vert \zeta\right\rangle $ satisfy the
Schr\"{o}dinger's equation (\ref{4}).

Substituting the states (\ref{16}) into (\ref{4}), we obtain the following
equation for $\Phi$:
\begin{equation}
\frac{\dot{\Phi}}{\Phi}=\frac{1}{2}\frac{\left\langle \zeta\left\vert \hat
{a}^{\dagger2}\right\vert \zeta\right\rangle }{\left\langle \zeta
|\zeta\right\rangle }\dot{\zeta}-\frac{i}{\hbar}\frac{\left\langle
\zeta\left\vert \hat{H}\right\vert \zeta\right\rangle }{\left\langle
\zeta|\zeta\right\rangle }. \label{17}%
\end{equation}
Using the representation (\ref{8}) together with the condition (\ref{15}), one
can easily calculate the mean values in (\ref{17}), with $\dot{\zeta}$ given
in (\ref{11}). Thus, the general solution from (\ref{17}), reads
\begin{equation}
\Phi=\frac{C}{f^{\varepsilon}}\exp\left(  -i\int\delta dt\right)  , \label{18}%
\end{equation}
where $C$ is a real normalization constant. Taking into account the
normalization condition, we find that:
\begin{equation}
\left\langle \zeta|\zeta\right\rangle =1\Rightarrow C=\mu^{\varepsilon/2}.
\label{18a}%
\end{equation}

Then, the normalized states $\left\vert \zeta\right\rangle $ that satisfy the
Schr\"{o}dinger's equation are given by
\begin{align}
\left\vert \zeta\right\rangle  &  =\frac{\sqrt{\mu^{\varepsilon}}%
}{f^{\varepsilon}}\exp\left(  -i\int\delta dt\right)  \exp\left(  -\frac{1}%
{2}\zeta\hat{a}^{\dagger2}\right)  \left\vert 0,\varepsilon\right\rangle
\nonumber\\
&  =\frac{\sqrt{\mu^{\varepsilon}}}{f^{\varepsilon}}\exp\left(  -i\int\delta
dt\right)
{\displaystyle\sum\limits_{n=0}^{\infty}}
\left(  -\zeta\right)  ^{n}\sqrt{\frac{\Gamma\left(  n+\varepsilon\right)
}{n!\Gamma\left(  \varepsilon\right)  }}\left\vert 2n,\varepsilon\right\rangle
. \label{19}%
\end{align}
The above sum converges as long as the condition $\left\vert \zeta\right\vert
<1$ is satisfied.

On the other hand, we can express $f$ and $\mu$ in terms of squeeze parameter
in the form:
\begin{equation}
f=f_{0}\exp\left[  -i\int\left(  \alpha^{\ast}\zeta-\beta\right)  dt\right]
,\text{ \ }\mu=\left\vert f\right\vert ^{2}\left(  1-\left\vert \zeta
\right\vert ^{2}\right)  . \label{19a}%
\end{equation}
From here, the states $\left\vert \zeta\right\rangle $ take the form
\begin{equation}
\left\vert \zeta\right\rangle =\left(  1-\left\vert \zeta\right\vert
^{2}\right)  ^{\frac{\varepsilon}{2}}e^{i\vartheta}%
{\displaystyle\sum\limits_{n=0}^{\infty}}
\left(  -\zeta\right)  ^{n}\sqrt{\frac{\Gamma\left(  n+\varepsilon\right)
}{n!\Gamma\left(  \varepsilon\right)  }}\left\vert 2n,\varepsilon\right\rangle
, \label{23a}%
\end{equation}
where the phase $\vartheta$ is given by
\begin{equation}
\vartheta=\int\left[  \varepsilon\operatorname{Re}\left(  \alpha\zeta^{\ast
}\right)  -\varepsilon\beta-\delta\right]  dt. \label{23b}%
\end{equation}
In what follows, we call the time-dependent states (\ref{23a}) generalized
SVS. Notice that, if we assume $\varepsilon=1/2$ and $\zeta=e^{i\theta}%
\tanh\left(  r\right)  \Longrightarrow\mu=1$, the Eq. (\ref{23a}) takes the
form:
\begin{equation}
\left\vert \zeta\right\rangle =\frac{e^{i\vartheta}}{\sqrt{\cosh(r)}}%
\sum_{n=0}^{\infty}\frac{\sqrt{(2n)!}}{2^{n}n!}\left[  -e^{i\theta}%
\tanh(r)\right]  ^{n}\left\vert 2n,\frac{1}{2}\right\rangle , \label{23c}%
\end{equation}
where $\Gamma\left(  n+\frac{1}{2}\right)  =\frac{\sqrt{\pi}4^{-n}(2n)!}{n!}$,
$\Gamma\left(  1/2\right)  =\sqrt{\pi}$. So, we can conclude that the states
obtained in Eq. (\ref{23c}) reproduce the usual case \cite{QUE2001}. In this
case, the Eq. (\ref{23a}) corresponds to a generalization of the SVS,
differing by a time-dependent phase factor due to the time evolution that was
included in our analysis. We must highlight that when identifying
$\varepsilon=2k$, the states (\ref{23a}) coincide with the time-independent CS
of the $SU\left(  1,1\right)  $ group constructed by Peremolov \cite{PER1986}.

\begin{figure}[ptb]
\subfloat[]{\includegraphics[width=8cm,height=6cm]{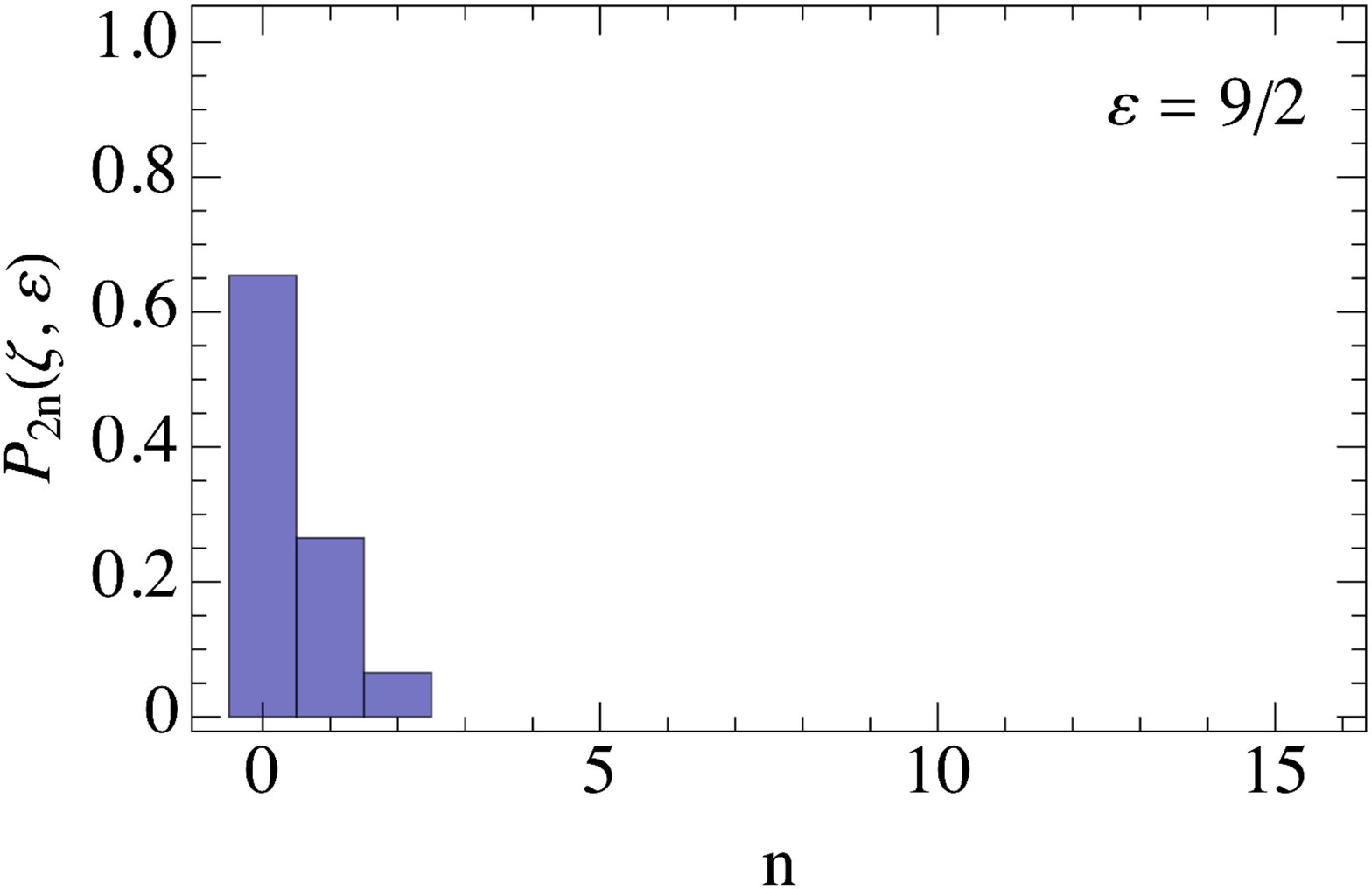}
}\hfill{}\subfloat[]{\includegraphics[width=8cm,height=6cm]{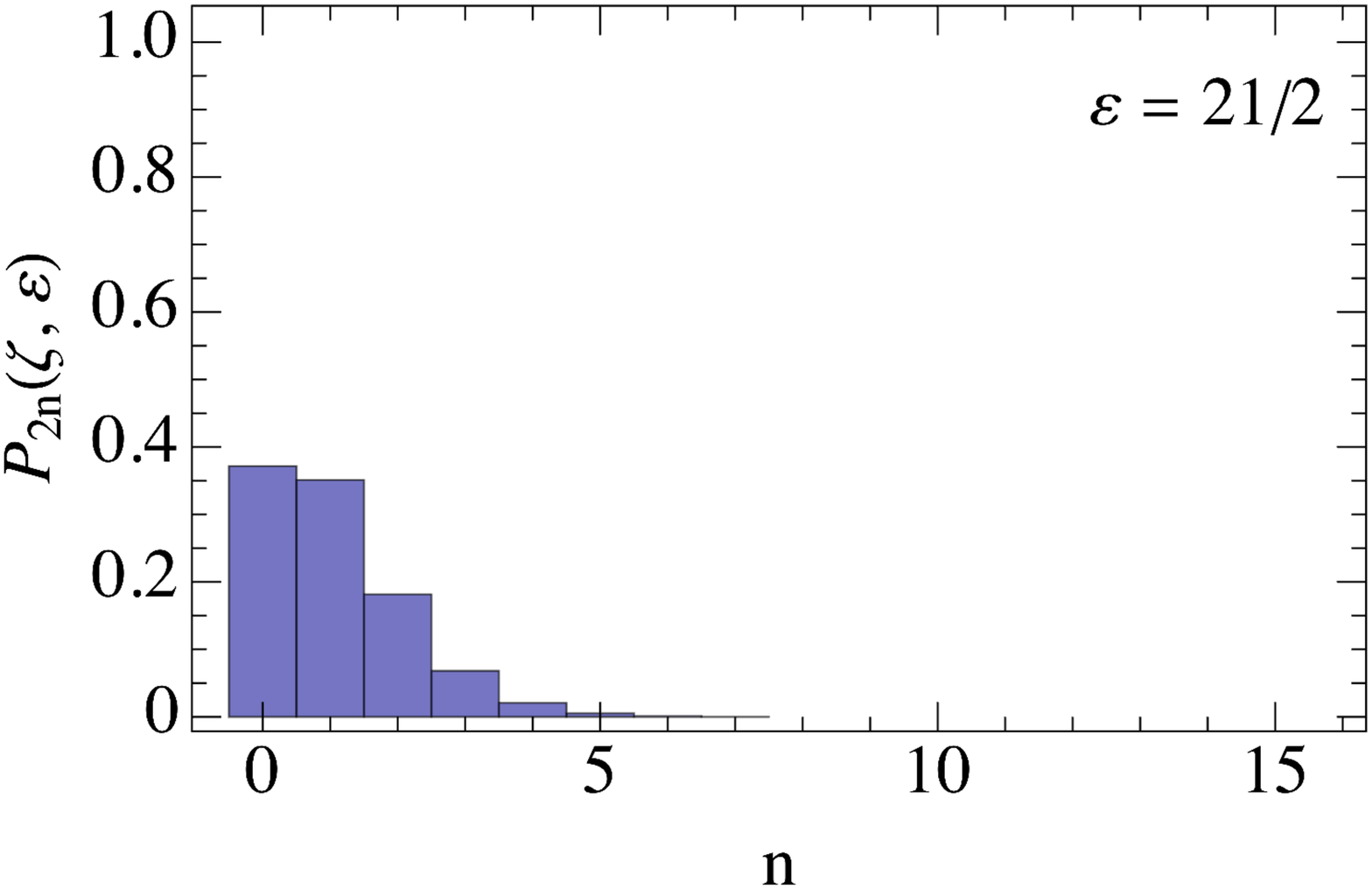}
}\vfill{} \subfloat[]{\includegraphics[width=8cm,height=6cm]{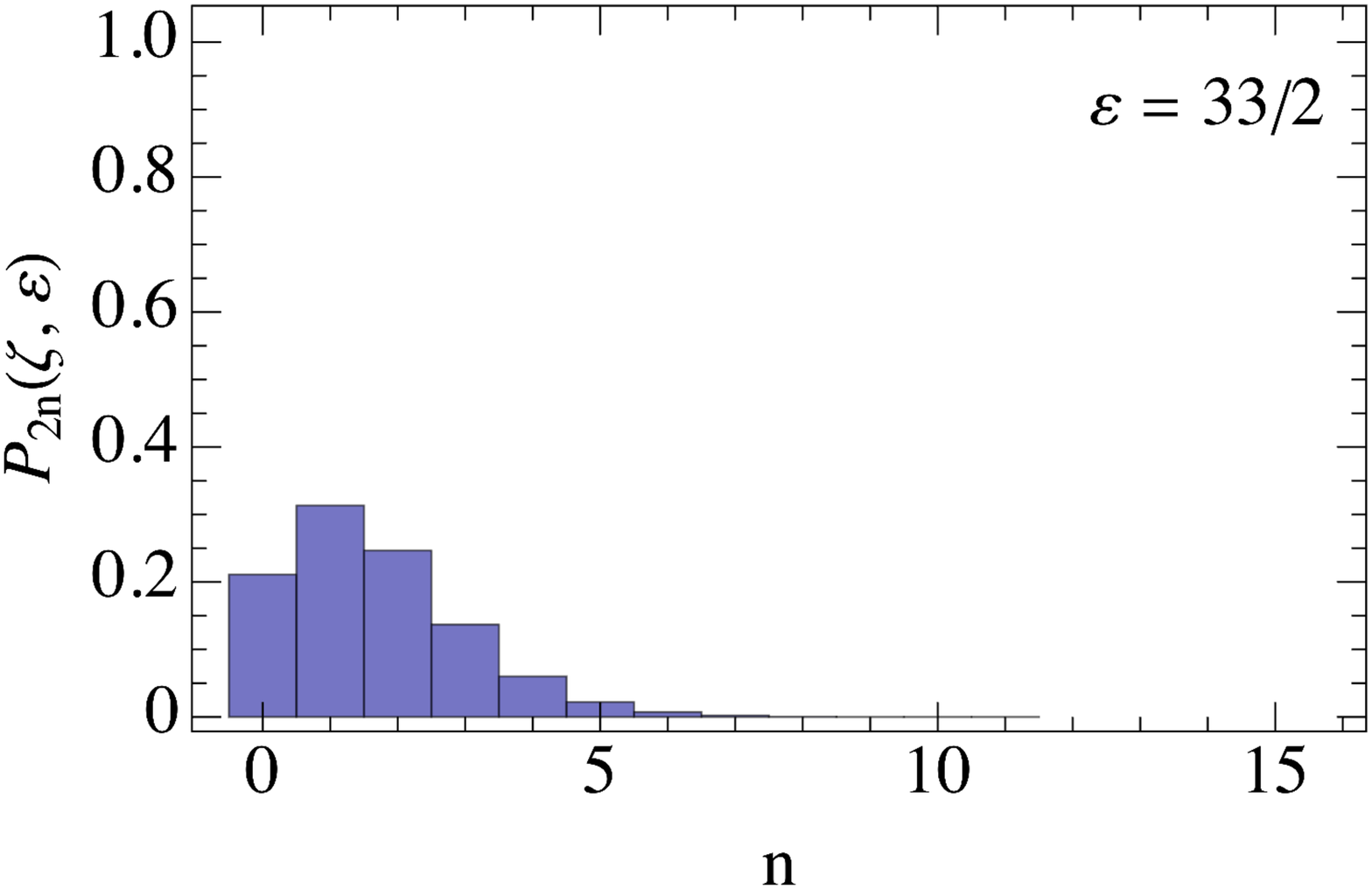}
}\hfill{}\subfloat[]{\includegraphics[width=8cm,height=6cm]{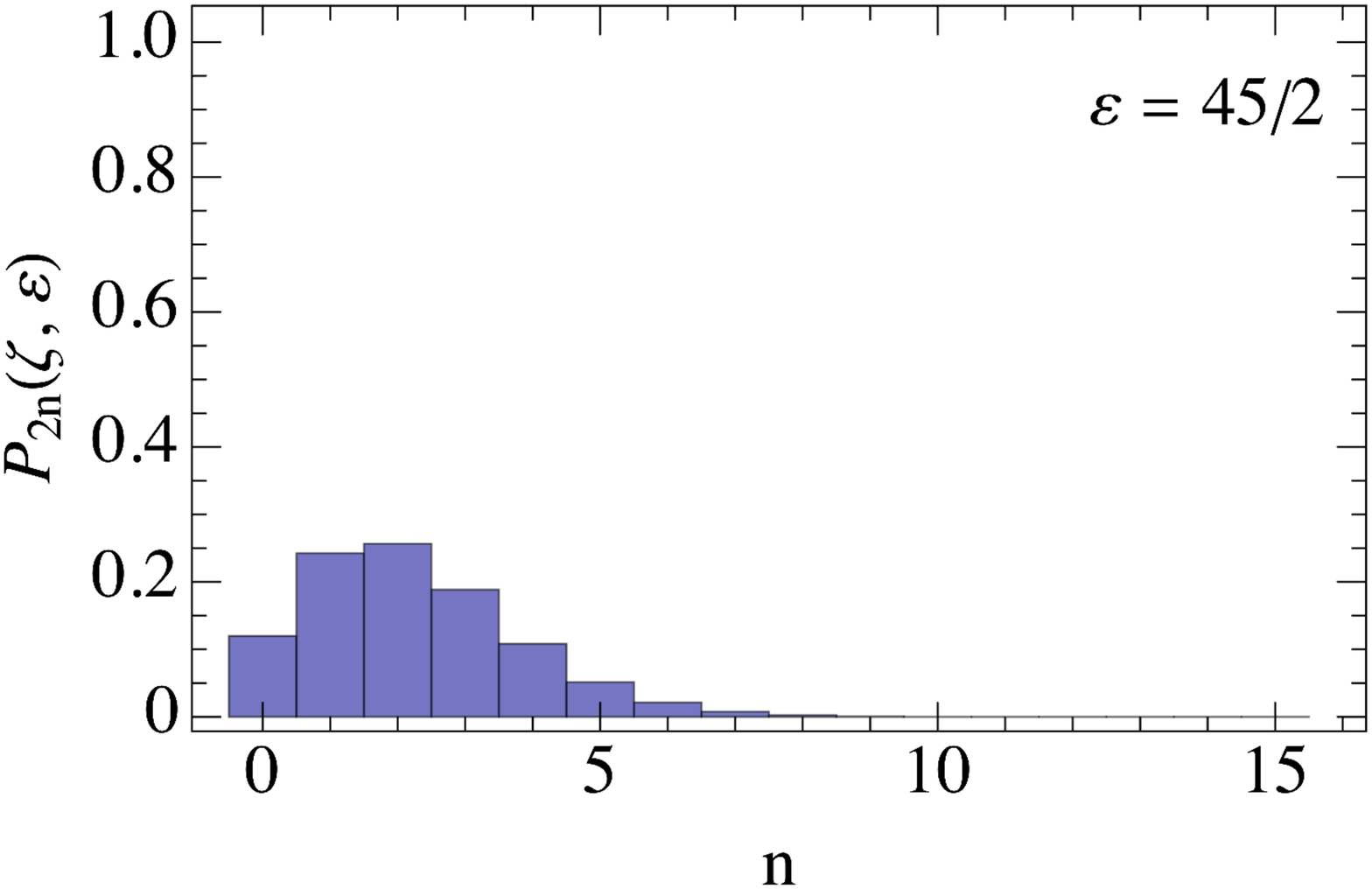}
}\caption{Probability transition of the generalized SVS by considering fixed
value for $\left|  \zeta\right|  =0.3$.}%
\label{fig1}%
\end{figure}

The overlap of two generalized SVS with different $\zeta$, for example
$\left\langle \zeta_{1}|\zeta_{2}\right\rangle $ reads
\begin{equation}
\left\langle \zeta_{1}|\zeta_{2}\right\rangle =\frac{\left(  1-\left\vert
\zeta_{1}\right\vert ^{2}\right)  ^{\frac{\varepsilon}{2}}\left(  1-\left\vert
\zeta_{2}\right\vert ^{2}\right)  ^{\frac{\varepsilon}{2}}}{\left(
1-\zeta_{1}^{\ast}\zeta_{2}\right)  ^{\varepsilon}}\exp\left\{  i\varepsilon
\int\operatorname{Re}\left[  \alpha\left(  \zeta_{2}^{\ast}-\zeta_{1}^{\ast
}\right)  \right]  dt\right\}  . \label{20}%
\end{equation}
In turn, the probability transition $P_{2n}\left(  \zeta,\varepsilon\right)
=\left\vert \left\langle \varepsilon,2n|\zeta\right\rangle \right\vert ^{2}$
is given by
\begin{equation}
P_{2n}\left(  \zeta,\varepsilon\right)  =\frac{\left(  1-\left\vert
\zeta\right\vert ^{2}\right)  ^{\varepsilon}}{\Gamma\left(  \varepsilon
\right)  }\frac{\Gamma\left(  n+\varepsilon\right)  \left\vert \zeta
\right\vert ^{2n}}{n!}. \label{21}%
\end{equation}
The probability transition (\ref{21}) has been plotted in Fig. \ref{fig1}. As
we see, the increase in the value of the $\varepsilon$ parameter implies a
larger dispersion of the probability transition.

\subsection{Completeness relation on squeeze parameter}

Considerations on the completeness relation for the SVS and squeezed odd
number states, in the context of canonical algebra, can be seen in Refs.
\citep{CHA1996,QUE2001}. Here, our contribution is to obtain the completeness
relation in the context of the WHA, and as will be seen, its existence depends
on the $\varepsilon$-parameter.

Let us consider a weight function $w\left(  \zeta\right)  $ such that the
states $\left\vert \zeta\right\rangle $ lead to the following closure
relation
\begin{equation}
\int_{%
\mathbb{C}%
}\left\vert \zeta\right\rangle \left\langle \zeta\right\vert w\left(
\zeta\right)  d^{2}\zeta=1. \label{R1}%
\end{equation}

In turn, substituting the states (\ref{23a}) into Eq. (\ref{R1}), we find
\begin{equation}%
{\displaystyle\sum\limits_{n,m=0}^{\infty}}
\left\vert 2n,\varepsilon\right\rangle \left\langle \varepsilon,2m\right\vert
\sqrt{\frac{\Gamma\left(  n+\varepsilon\right)  }{n!\Gamma\left(
\varepsilon\right)  }\frac{\Gamma\left(  m+\varepsilon\right)  }%
{m!\Gamma\left(  \varepsilon\right)  }}\int_{%
\mathbb{C}%
}\left(  1-\left\vert \zeta\right\vert ^{2}\right)  ^{\varepsilon}\left(
-\zeta\right)  ^{n}\left(  -\zeta^{\ast}\right)  ^{m}w\left(  \zeta\right)
d^{2}\zeta=1. \label{R2}%
\end{equation}
Now, by using polar coordinates in the above relation,
\begin{equation}
\zeta=r_{\zeta}e^{i\theta_{\zeta}},\text{ \ }d^{2}\zeta=r_{\zeta}dr_{\zeta
}d\theta_{\zeta},\text{ \ }0\leq r_{\zeta}<1,\text{ \ }0\leq\theta_{\zeta}%
\leq2\pi, \label{R3}%
\end{equation}
we get
\begin{align}
&
{\displaystyle\sum\limits_{n,m=0}^{\infty}}
\left\vert 2n,\varepsilon\right\rangle \left\langle \varepsilon,2m\right\vert
\sqrt{\frac{\Gamma\left(  n+\varepsilon\right)  }{n!\Gamma\left(
\varepsilon\right)  }\frac{\Gamma\left(  m+\varepsilon\right)  }%
{m!\Gamma\left(  \varepsilon\right)  }}\int_{0}^{1}\int_{0}^{2\pi}\left(
1-r_{\zeta}^{2}\right)  ^{\varepsilon}\left(  -r_{\zeta}\right)  ^{n}\left(
-r_{\zeta}\right)  ^{m}\times\nonumber\\
&  w\left(  \zeta\right)  \exp\left[  i\left(  n-m\right)  \theta_{\zeta
}\right]  r_{\zeta}dr_{\zeta}d\theta_{\zeta}=1. \label{R4}%
\end{align}

\begin{figure}[ptb]
\includegraphics[scale=0.4]{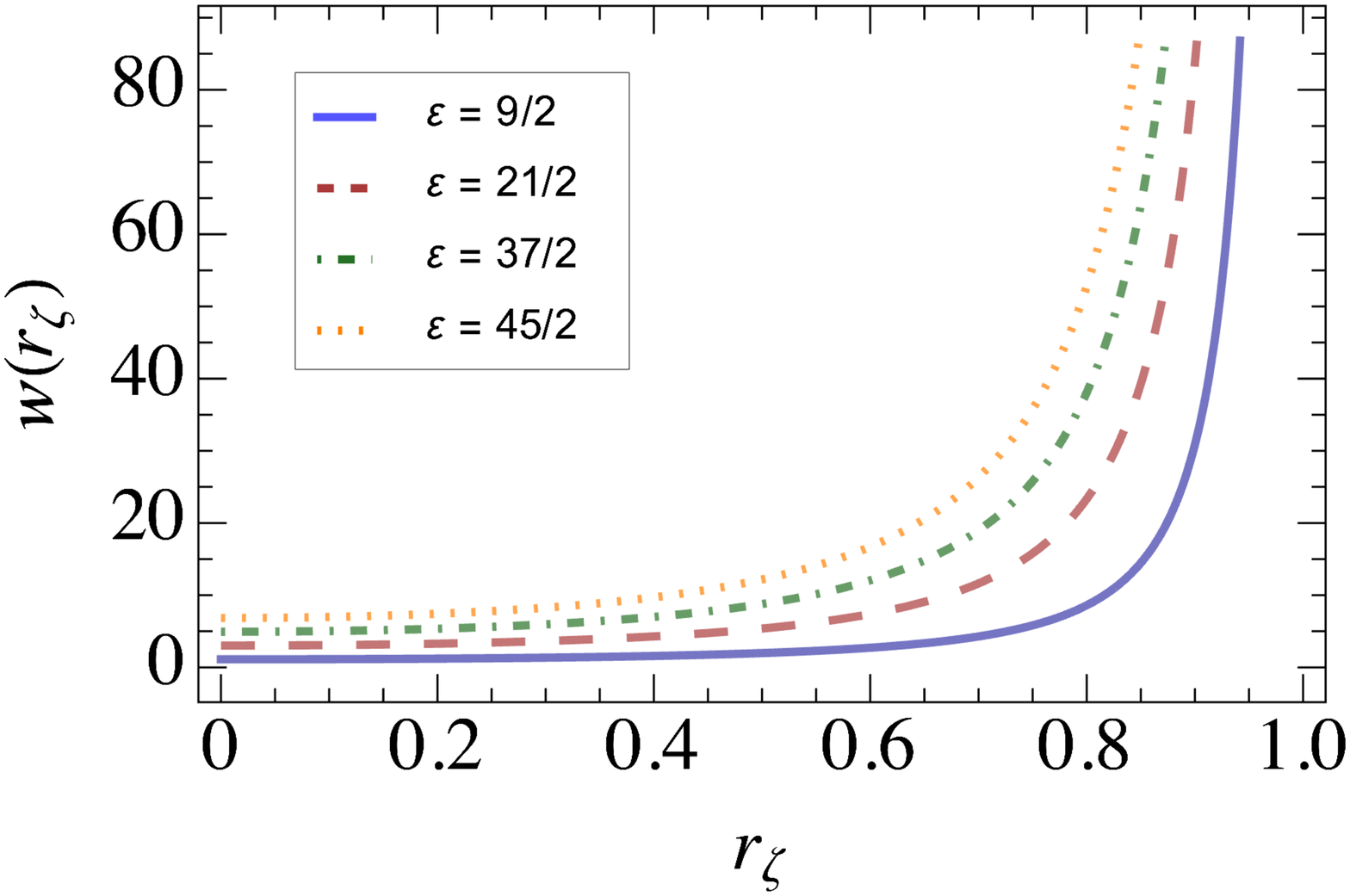}\caption{Graphics of $w\left(  r_{\zeta
}\right)  $.}%
\label{figw}%
\end{figure}

On the other hand, we may readily show that $w\left(  \zeta\right)  =w\left(
r_{\zeta}\right)  $, which implies that the integral on $\theta_{\zeta}$
becomes
\begin{equation}
\int_{0}^{2\pi}d\theta_{\zeta}\exp\left[  i\left(  n-m\right)  \theta_{\zeta
}\right]  =2\pi\delta_{n,m}. \label{R5}%
\end{equation}
From here, we can write (\ref{R4}) in the form
\begin{equation}
2\pi%
{\displaystyle\sum\limits_{n=0}^{\infty}}
\left\vert 2n,\varepsilon\right\rangle \left\langle \varepsilon,2n\right\vert
\frac{\Gamma\left(  n+\varepsilon\right)  }{n!\Gamma\left(  \varepsilon
\right)  }\int_{0}^{1}\left(  1-r_{\zeta}^{2}\right)  ^{\varepsilon}r_{\zeta
}^{2n+1}w\left(  r_{\zeta}\right)  dr_{\zeta}=1. \label{R6}%
\end{equation}
By using the following relationship among gamma functions
\begin{equation}
\frac{\Gamma\left(  x\right)  \Gamma\left(  y\right)  }{\Gamma\left(
x+y\right)  }=2\int_{0}^{1}\left(  1-t^{2}\right)  ^{y-1}t^{2x-1}dt,\text{
\ }\operatorname{Re}\left(  x\right)  >0,\text{ \ }\operatorname{Re}\left(
y\right)  >0, \label{R9}%
\end{equation}
one can see that the weight function $w\left(  r_{\zeta}\right)  $ must have
the form
\begin{equation}
w\left(  r_{\zeta}\right)  =\frac{\Gamma\left(  \varepsilon\right)  }%
{\pi\Gamma\left(  \varepsilon-1\right)  \left(  1-r_{\zeta}^{2}\right)  ^{2}%
}=\frac{\varepsilon-1}{\pi\left(  1-r_{\zeta}^{2}\right)  ^{2}},\text{
\ }\varepsilon>1, \label{R10}%
\end{equation}
in order to ensure that the completeness relation (\ref{R1}) is satisfied. It
is important to highlight that the condition $\varepsilon>1$ leads to a
positive weight function $w\left(  r_{\zeta}\right)  $. At the same time, this
condition shows that the canonical algebra ($\varepsilon=1/2$) does not allow
to obtain a completeness relation for the generalized SVS. It follows from
(\ref{20}) and (\ref{R1}) that these states form an overcomplete set of states
on the Hilbert space. The plot of weight function can be seen in the Fig.
\ref{figw}.

\section{Time-dependent generalized CS\label{IV}}

In Ref. \citep{PER2021}, the SCS has been built through the nonunitary
approach by considering the canonical algebra. In another way, by considering
the WHA, it is not possible to obtain the SCS since, from
Baker--Campbell--Hausdorff theorem, we cannot establish a direct relationship
between the para-Bose operators and integral of motion, due the deformed
commutation relation. However, instead of following the nonunitary approach,
we will construct the eigenstates of the integral of motion $\hat{A}$, in the
form
\begin{equation}
\hat{A}\left\vert z,t\right\rangle =z\left\vert z,t\right\rangle , \label{22}%
\end{equation}
where $z$ is a complex constant. The Eq. (\ref{22}) allows us to assume
\begin{equation}
\varphi_{0}=0\Longrightarrow u=0, \label{23}%
\end{equation}
without loss of generality.

The states $\left\vert z,t\right\rangle $ can be expanded in terms of the
para-Bose number states, as follows
\begin{equation}
\left\vert z,t\right\rangle =%
{\displaystyle\sum\limits_{n=0}^{\infty}}
c_{n}\left\vert n,\varepsilon\right\rangle =%
{\displaystyle\sum\limits_{n=0}^{\infty}}
\left(  c_{2n}\left\vert 2n,\varepsilon\right\rangle +c_{2n+1}\left\vert
2n+1,\varepsilon\right\rangle \right)  , \label{24}%
\end{equation}
where $c_{n}=c_{n}\left(  t\right)  $ are time-dependent coefficients and will
be determined such that the Eqs. (\ref{22}) and (\ref{4}) will be satisfied.
Substituting (\ref{5}), (\ref{9}) and (\ref{24}) into (\ref{22}), we find the
following equations for the coefficients
\begin{align}
&  c_{2n}=\sqrt{\frac{n!\Gamma\left(  \varepsilon\right)  }{\Gamma\left(
n+\varepsilon\right)  }}\left(  -\frac{g}{f}\right)  ^{n}L_{n}^{\varepsilon
-1}\left(  \frac{z^{2}}{2gf}\right)  c_{0},\text{ \ }\nonumber\\
&  c_{2n+1}=\frac{z}{f}\sqrt{\frac{n!\Gamma\left(  \varepsilon\right)
}{2\Gamma\left(  n+\varepsilon+1\right)  }}\left(  -\frac{g}{f}\right)
^{n}L_{n}^{\varepsilon}\left(  \frac{z^{2}}{2gf}\right)  c_{0}, \label{25}%
\end{align}
where $L_{n}^{\alpha}\left(  x\right)  $ are the associated Laguerre
polynomials and $c_{0}$ is a time-dependent function, which will be determined
such that the states $\left\vert z,t\right\rangle $ satisfy the
Schr\"{o}dinger's equation (\ref{4}).

Here, it is convenient to introduce the squeeze $\zeta=g/f$ and displacement
$\xi=z/f$ parameters to rewrite the states $\left\vert z,t\right\rangle
\longrightarrow\left\vert \zeta,\xi\right\rangle $, in the form
\begin{equation}
\left\vert \zeta,\xi\right\rangle =\sqrt{\Gamma\left(  \varepsilon\right)
}c_{0}%
{\displaystyle\sum\limits_{n=0}^{\infty}}
\left(  -\zeta\right)  ^{n}\sqrt{n!}\left[  \frac{L_{n}^{\varepsilon-1}\left(
\frac{\xi^{2}}{2\zeta}\right)  }{\sqrt{\Gamma\left(  n+\varepsilon\right)  }%
}\left\vert 2n,\varepsilon\right\rangle +\frac{\xi L_{n}^{\varepsilon}\left(
\frac{\xi^{2}}{2\zeta}\right)  }{\sqrt{2\Gamma\left(  n+\varepsilon+1\right)
}}\left\vert 2n+1,\varepsilon\right\rangle \right]  , \label{25a}%
\end{equation}
with $\zeta$ and $\xi$ satisfy the following differential equations:
\begin{equation}
\dot{\zeta}=i\alpha^{\ast}\zeta^{2}-2i\beta\zeta+i\alpha,\text{ \ }\dot{\xi
}=i\left(  \alpha^{\ast}\zeta-\beta\right)  \xi. \label{25b}%
\end{equation}
From normalization condition
\begin{equation}
\left\langle \xi,\zeta|\zeta,\xi\right\rangle =1, \label{26}%
\end{equation}
we found the following form for the function $c_{0}$,
\begin{equation}
c_{0}=\left(  \frac{\xi}{\sqrt{2}}\right)  ^{\varepsilon-1}\sqrt
{\frac{1-\left\vert \zeta\right\vert ^{2}}{\Gamma\left(  \varepsilon\right)
}}\frac{\exp\left\{  \frac{\zeta^{\ast}\xi^{2}}{2\left(  1-\left\vert
\zeta\right\vert ^{2}\right)  }+i\int\left[  \operatorname{Re}\left(
\alpha\zeta^{\ast}\right)  -\beta\right]  dt+i\phi\right\}  }{\sqrt
{I_{\varepsilon-1}\left(  \frac{\left\vert \xi\right\vert ^{2}}{1-\left\vert
\zeta\right\vert ^{2}}\right)  +I_{\varepsilon}\left(  \frac{\left\vert
\xi\right\vert ^{2}}{1-\left\vert \zeta\right\vert ^{2}}\right)  }},
\label{27}%
\end{equation}
where $I_{\kappa}\left(  Z\right)  $ is the modified Bessel function of the
first kind, and $\phi$ is a time-dependent real function. Thus, the normalized
states $\left\vert \zeta,\xi\right\rangle $ take the form
\begin{align}
\left\vert \zeta,\xi\right\rangle  &  =\left(  \frac{\xi}{\sqrt{2}}\right)
^{\varepsilon-1}\sqrt{\frac{1-\left\vert \zeta\right\vert ^{2}}{I_{\varepsilon
-1}\left(  \frac{\left\vert \xi\right\vert ^{2}}{1-\left\vert \zeta\right\vert
^{2}}\right)  +I_{\varepsilon}\left(  \frac{\left\vert \xi\right\vert ^{2}%
}{1-\left\vert \zeta\right\vert ^{2}}\right)  }}\exp\left\{  \frac{\zeta
^{\ast}\xi^{2}}{2\left(  1-\left\vert \zeta\right\vert ^{2}\right)  }%
+i\int\left[  \operatorname{Re}\left(  \alpha\zeta^{\ast}\right)
-\beta\right]  dt+i\phi\right\} \nonumber\\
&  \times%
{\displaystyle\sum\limits_{n=0}^{\infty}}
\left(  -\zeta\right)  ^{n}\sqrt{n!}\left[  \frac{L_{n}^{\varepsilon-1}\left(
\frac{\xi^{2}}{2\zeta}\right)  }{\sqrt{\Gamma\left(  n+\varepsilon\right)  }%
}\left\vert 2n,\varepsilon\right\rangle +\frac{\xi L_{n}^{\varepsilon}\left(
\frac{\xi^{2}}{2\zeta}\right)  }{\sqrt{2\Gamma\left(  n+\varepsilon+1\right)
}}\left\vert 2n+1,\varepsilon\right\rangle \right]  , \label{28}%
\end{align}

Substituting $\left\vert \zeta,\xi\right\rangle $ into Schr\"{o}dinger's
equation, we find the following expression for $\phi$,
\begin{equation}
\phi=-\int\delta dt. \label{29}%
\end{equation}
Therefore, the normalized states $\left\vert \zeta,\xi\right\rangle $ that
satisfy the Schr\"{o}dinger's equation are given by
\begin{align}
\left\vert \zeta,\xi\right\rangle  &  =\left(  \frac{\xi}{\sqrt{2}}\right)
^{\varepsilon-1}\sqrt{\frac{1-\left\vert \zeta\right\vert ^{2}}{I_{\varepsilon
-1}\left(  \frac{\left\vert \xi\right\vert ^{2}}{1-\left\vert \zeta\right\vert
^{2}}\right)  +I_{\varepsilon}\left(  \frac{\left\vert \xi\right\vert ^{2}%
}{1-\left\vert \zeta\right\vert ^{2}}\right)  }}\exp\left[  \frac{\zeta^{\ast
}\xi^{2}}{2\left(  1-\left\vert \zeta\right\vert ^{2}\right)  }+i\tilde
{\vartheta}\right] \nonumber\\
&  \times%
{\displaystyle\sum\limits_{n=0}^{\infty}}
\left(  -\zeta\right)  ^{n}\sqrt{n!}\left[  \frac{L_{n}^{\varepsilon-1}\left(
\frac{\xi^{2}}{2\zeta}\right)  }{\sqrt{\Gamma\left(  n+\varepsilon\right)  }%
}\left\vert 2n,\varepsilon\right\rangle +\frac{\xi L_{n}^{\varepsilon}\left(
\frac{\xi^{2}}{2\zeta}\right)  }{\sqrt{2\Gamma\left(  n+\varepsilon+1\right)
}}\left\vert 2n+1,\varepsilon\right\rangle \right]  , \label{30}%
\end{align}
where
\begin{equation}
\tilde{\vartheta}=\int\left[  \operatorname{Re}\left(  \alpha\zeta^{\ast
}\right)  -\beta-\delta\right]  dt. \label{30.1}%
\end{equation}
In the following, we call the time-dependent states (\ref{30}) generalized
para-Bose CS.

It's worth mentioning that the states (\ref{30}) present a set of information
that, to our knowledge, has not been previously derived, see, e.g.,
\cite{SHA1978,SHA1981,BIS1980,SAX1986,MOJ2018a,MOJ2018b,Ald2017,Bag1997,DEH2015,ALD2021,ALD2017}%
. Such information consists of the explicit form of the squeeze, displacement,
and deformation parameters (Wigner-parameter). These parameters are present
explicitly in the standard deviation and probability transition. Furthermore,
the time-dependent states (\ref{30}) have been expanded in terms of the
time-independent para-Bose number states. Thus, the states (\ref{30}) lead to
a deeper understanding of the physical systems described by the Hamiltonian
(\ref{3}). As we will see from section \ref{V}, the Hamiltonian can be seen as
a generalization of Calogero model \cite{CAL1971}.

In particular, taking into account the condition $\zeta=0\Longrightarrow
\alpha=0$, the states (\ref{30}) take the form
\begin{align}
\left\vert \xi\right\rangle  &  =\left(  \frac{\xi}{\sqrt{2}}\right)
^{\varepsilon-1}\frac{\exp\left[  -i\int\left(  \beta+\delta\right)
dt\right]  }{\sqrt{I_{\varepsilon-1}\left(  \left\vert \xi\right\vert
^{2}\right)  +I_{\varepsilon}\left(  \left\vert \xi\right\vert ^{2}\right)  }}%
{\displaystyle\sum\limits_{n=0}^{\infty}}
\left(  \frac{\xi^{2}}{2}\right)  ^{n}\left[  \frac{\left\vert 2n,\varepsilon
\right\rangle }{\sqrt{n!\Gamma\left(  n+\varepsilon\right)  }}+\frac
{\xi\left\vert 2n+1,\varepsilon\right\rangle }{\sqrt{2n!\Gamma\left(
n+\varepsilon+1\right)  }}\right] \nonumber\\
&  =\sqrt{\Gamma\left(  \varepsilon\right)  }\exp\left[  -i\int\left(
\beta+\delta\right)  dt\right]  \left(  \frac{\hat{a}^{\dagger}}{\sqrt{2}%
}\right)  ^{1-\varepsilon}\frac{I_{\varepsilon-1}\left(  \xi\hat{a}^{\dagger
}\right)  +I_{\varepsilon}\left(  \xi\hat{a}^{\dagger}\right)  }%
{\sqrt{I_{\varepsilon-1}\left(  \left\vert \xi\right\vert ^{2}\right)
+I_{\varepsilon}\left(  \left\vert \xi\right\vert ^{2}\right)  }}\left\vert
0,\varepsilon\right\rangle , \label{30.2}%
\end{align}
with $\dot{\xi}=-i\beta\xi$ and $\left\vert \xi\right\rangle =\left\vert
0,\xi\right\rangle $. Except for time evolution, these states correspond to
the para-Bose CS obtained in Ref. \cite{SHA1978}. In turn, we have that the
condition $\varepsilon=1/2$ reduces the states (\ref{30.2}) to the form:
\begin{equation}
\left\vert \xi\right\rangle =\exp\left(  \int\frac{\beta+2\delta}{2i}%
dt-\frac{\left\vert \xi\right\vert ^{2}}{2}\right)  \exp\left(  \xi\hat
{a}^{\dagger}\right)  \left\vert 0\right\rangle ,\text{ \ }\left\vert
0\right\rangle =\left\vert 0,\varepsilon=1/2\right\rangle , \label{30.3}%
\end{equation}
which are the time-dependent canonical CS \cite{GLA1966,MEH1966}.

The overlap of the states $\left\vert \zeta,\xi\right\rangle $ for different
squeeze $\zeta$ and displacement $\xi$ parameters is given by
\begin{align}
\left\langle \xi_{1},\zeta_{1}|\zeta,\xi\right\rangle  &  =\left(  \frac
{\xi\xi_{1}^{\ast}}{2}\right)  ^{\varepsilon-1}\sqrt{\frac{\left(
1-\left\vert \zeta\right\vert ^{2}\right)  \left(  1-\left\vert \zeta
_{1}\right\vert ^{2}\right)  }{\left[  I_{\varepsilon-1}\left(  \frac
{\left\vert \xi\right\vert ^{2}}{1-\left\vert \zeta\right\vert ^{2}}\right)
+I_{\varepsilon}\left(  \frac{\left\vert \xi\right\vert ^{2}}{1-\left\vert
\zeta\right\vert ^{2}}\right)  \right]  \left[  I_{\varepsilon-1}\left(
\frac{\left\vert \xi_{1}\right\vert ^{2}}{1-\left\vert \zeta_{1}\right\vert
^{2}}\right)  +I_{\varepsilon}\left(  \frac{\left\vert \xi_{1}\right\vert
^{2}}{1-\left\vert \zeta_{1}\right\vert ^{2}}\right)  \right]  }}\nonumber\\
&  \times\exp\left[  \frac{\zeta^{\ast}\xi^{2}\left(  1-\left\vert \zeta
_{1}\right\vert ^{2}\right)  +\zeta_{1}\xi_{1}^{\ast2}\left(  1-\left\vert
\zeta\right\vert ^{2}\right)  }{2\left(  1-\left\vert \zeta\right\vert
^{2}\right)  \left(  1-\left\vert \zeta_{1}\right\vert ^{2}\right)  }%
+i\int\operatorname{Re}\left(  \alpha\zeta^{\ast}-\alpha\zeta_{1}^{\ast
}\right)  dt\right] \nonumber\\
&  \times%
{\displaystyle\sum\limits_{n=0}^{\infty}}
\left(  \zeta\zeta_{1}^{\ast}\right)  ^{n}n!\left[  \frac{L_{n}^{\varepsilon
-1}\left(  \frac{\xi^{2}}{2\zeta}\right)  L_{n}^{\varepsilon-1}\left(
\frac{\xi_{1}^{\ast2}}{2\zeta_{1}^{\ast}}\right)  }{\Gamma\left(
n+\varepsilon\right)  }+\frac{\xi\xi_{1}^{\ast}}{2}\frac{L_{n}^{\varepsilon
}\left(  \frac{\xi^{2}}{2\zeta}\right)  L_{n}^{\varepsilon}\left(  \frac
{\xi_{1}^{\ast2}}{2\zeta_{1}^{\ast}}\right)  }{\Gamma\left(  n+\varepsilon
+1\right)  }\right]  . \label{30.4}%
\end{align}

Here, we can analyze the probability transition $P_{n}\left(  \zeta
,\xi,\varepsilon\right)  =\left\vert \left\langle \varepsilon,n|\zeta
,\xi\right\rangle \right\vert ^{2}$ of the number states $\left\vert
n,\varepsilon\right\rangle $ to the CS $\left\vert \zeta,\xi\right\rangle $:
\begin{align}
&  P_{n}\left(  \zeta,\xi,\varepsilon\right)  =\left(  \frac{\left\vert
\xi\right\vert ^{2}}{2}\right)  ^{\varepsilon-1}\frac{\left(  1-\left\vert
\zeta\right\vert ^{2}\right)  \exp\left[  \frac{\operatorname{Re}\left(
\zeta^{\ast}\xi^{2}\right)  }{1-\left\vert \zeta\right\vert ^{2}}\right]
}{I_{\varepsilon-1}\left(  \frac{\left\vert \xi\right\vert ^{2}}{1-\left\vert
\zeta\right\vert ^{2}}\right)  +I_{\varepsilon}\left(  \frac{\left\vert
\xi\right\vert ^{2}}{1-\left\vert \zeta\right\vert ^{2}}\right)  }m!\left\vert
\zeta\right\vert ^{2m}\left[  \frac{\left\vert L_{m}^{\varepsilon-1}\left(
\frac{\xi^{2}}{2\zeta}\right)  \right\vert ^{2}\delta_{n,2m}}{\Gamma\left(
m+\varepsilon\right)  }+\right. \nonumber\\
&  \left.  +\frac{\left\vert \xi\right\vert ^{2}\left\vert L_{m}^{\varepsilon
}\left(  \frac{\xi^{2}}{2\zeta}\right)  \right\vert ^{2}\delta_{n,2m+1}%
}{2\Gamma\left(  m+\varepsilon+1\right)  }\right]  \label{30a}%
\end{align}

\begin{figure}[ptb]
\subfloat[]{\includegraphics[width=8cm,height=6cm]{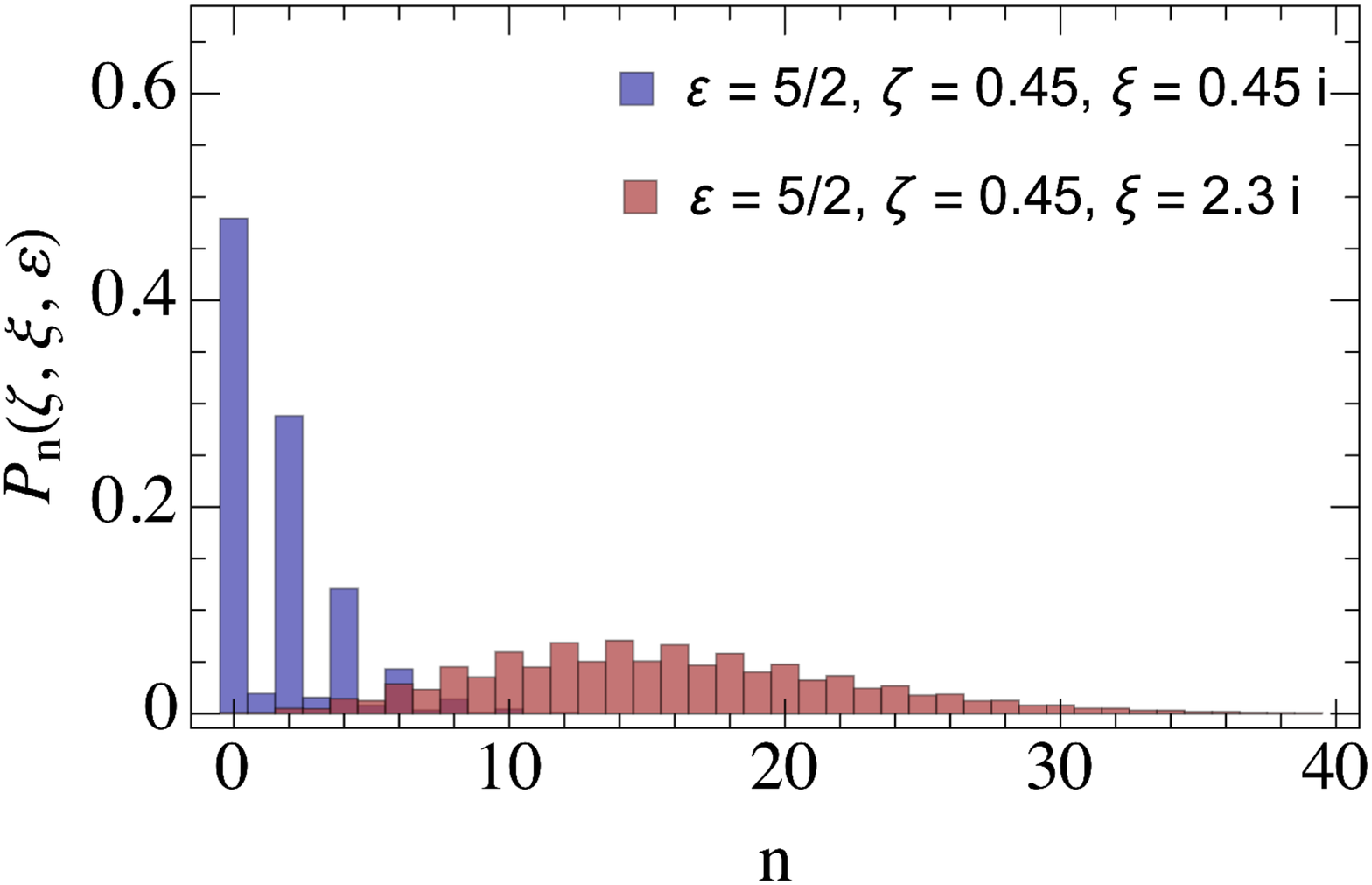}
}\hfill{}\subfloat[]{\includegraphics[width=8cm,height=6cm]{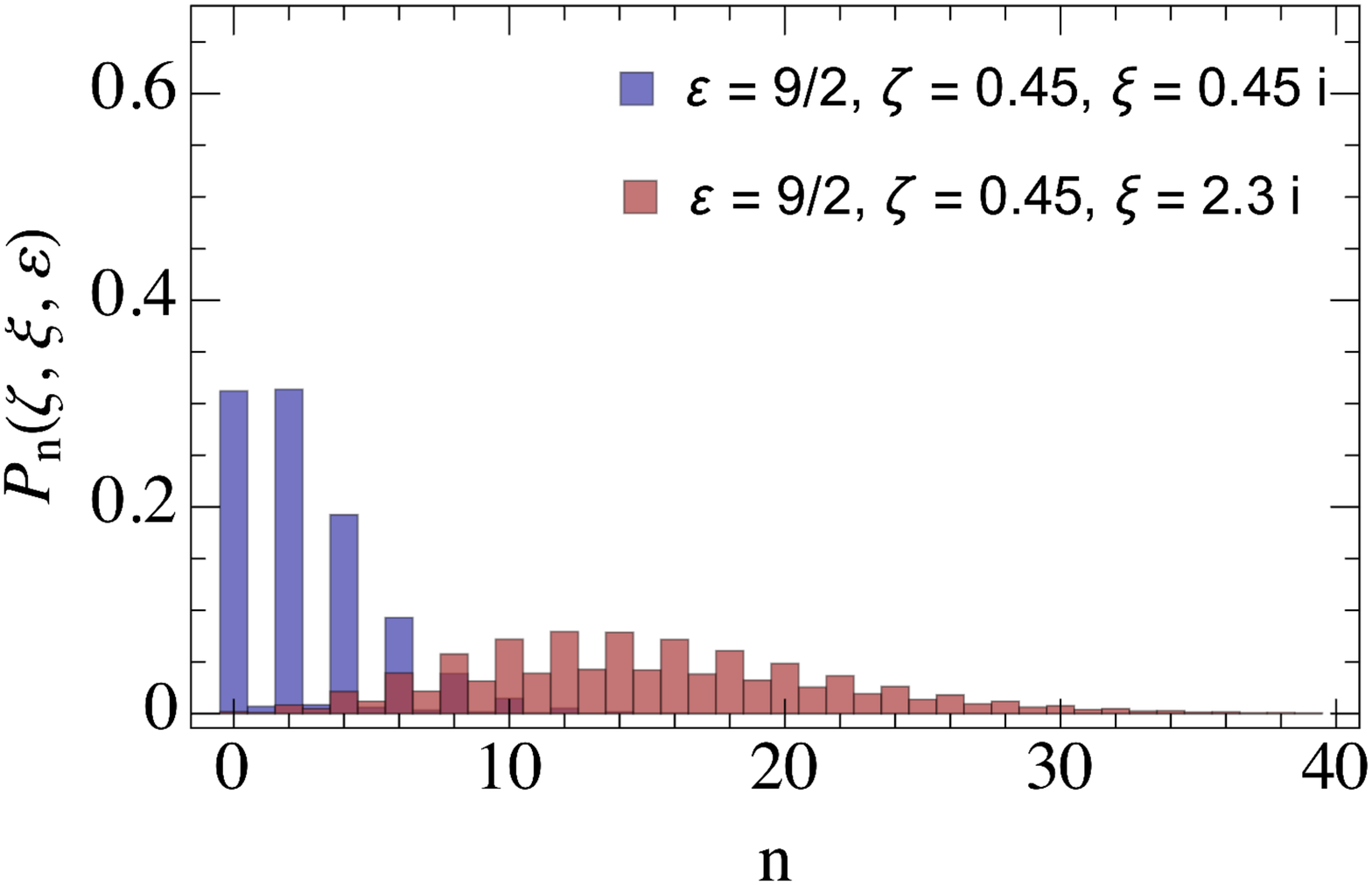}
}\vfill{} \subfloat[]{\includegraphics[width=8cm,height=6cm]{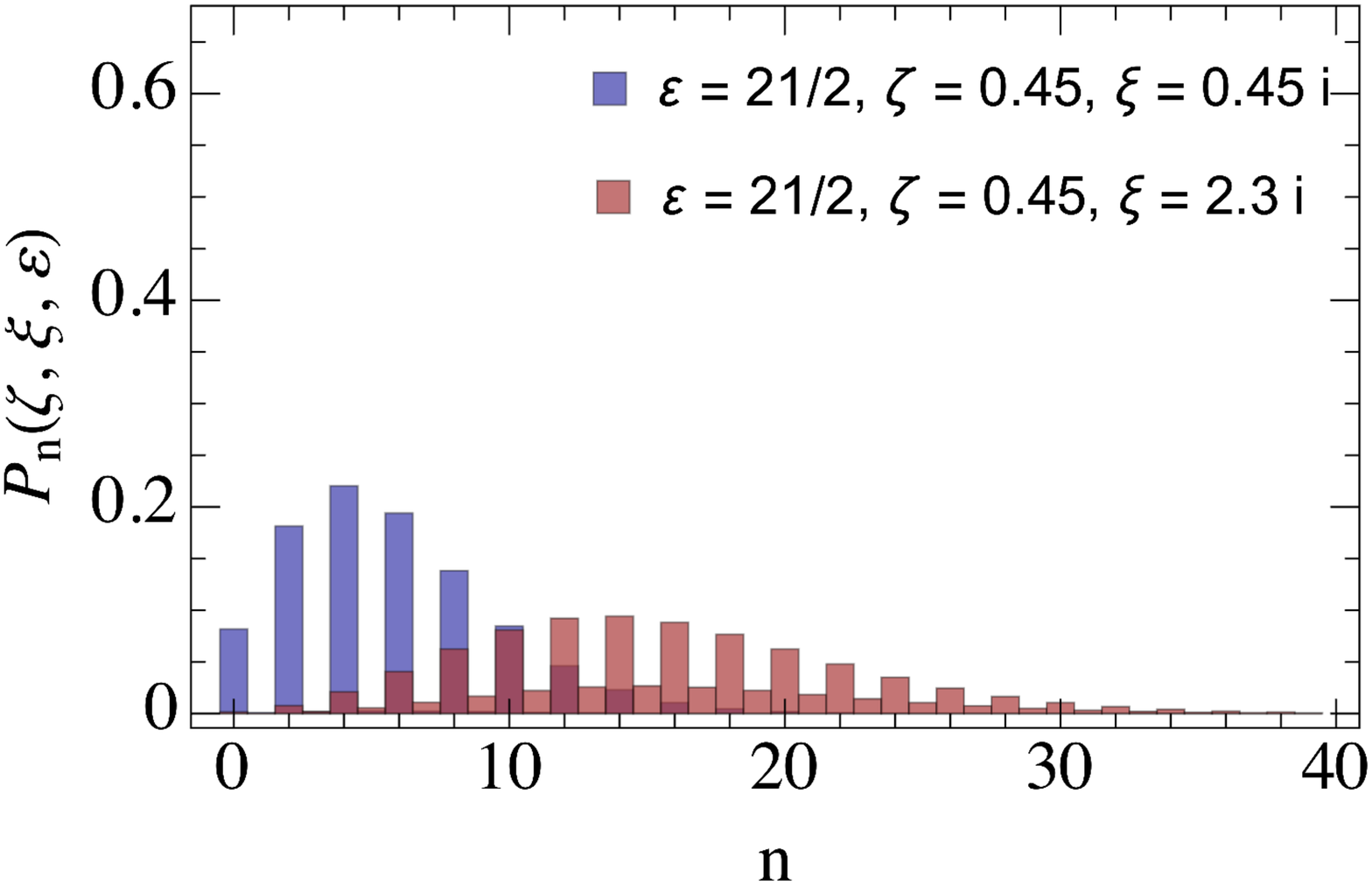}
}\hfill{}\subfloat[]{\includegraphics[width=8cm,height=6cm]{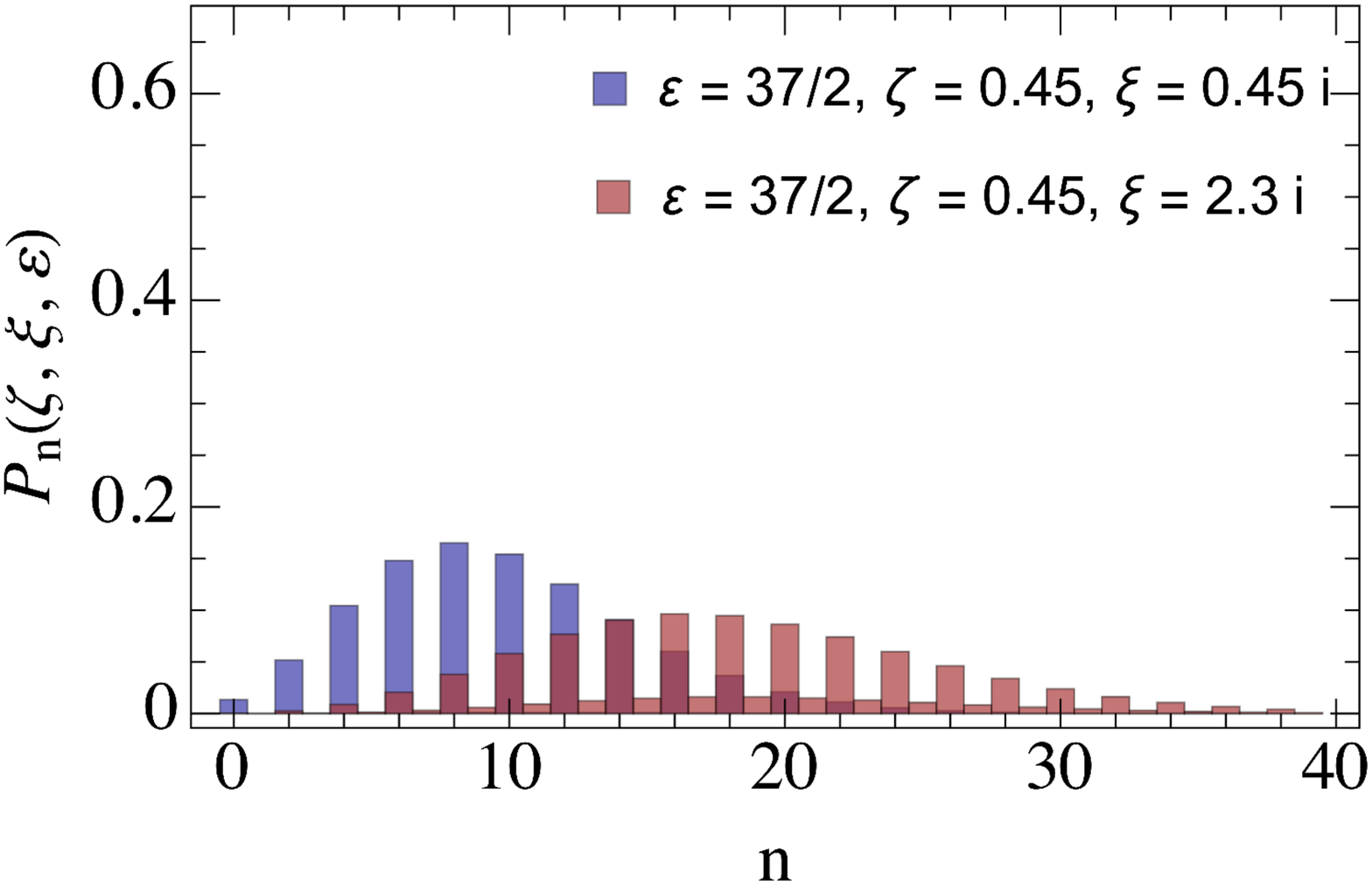}
}\caption{Probability transition of the generalized CS. }%
\label{fig3}%
\end{figure}

The probability transition $P_{n}\left(  \zeta,\xi,\varepsilon\right)  $ has
been shown in Fig. \ref{fig3}. As we can see, the $\xi$-parameter allows
access to odd states while the $\varepsilon$ controls the dispersion of the
probability density. Furthermore, higher values of the $\varepsilon$-parameter
allow access to states with higher principal quantum number $n$, and attenuate
the access to the odd states.

\subsection{Mean values and uncertainty relations\label{sec:Valor-esperado-e}}

Let us now consider the mean value of the momentum $\hat{P}$ and position
$\hat{x}$ operators, which satisfy the commutation relation
\begin{equation}
\left[  \hat{x},\hat{P}\right]  =i\hbar\left[  1+\left(  2\varepsilon
-1\right)  \hat{R}\right]  , \label{47}%
\end{equation}
in the generalized CS (\ref{30}). For this, we will rewrite these operators in
terms of the integrals of motion, as seen below
\begin{align}
&  \hat{a}=\frac{\hat{A}_{f}-\zeta\hat{A}_{f^{\ast}}^{\dagger}}{1-\left\vert
\zeta\right\vert ^{2}},\text{ \ }\hat{A}_{f}\equiv\frac{1}{f}\hat
{A},\nonumber\\
&  \hat{x}=\frac{\hat{a}+\hat{a}^{\dagger}}{\sqrt{2}}l=\frac{l}{\sqrt{2}}%
\frac{\left(  1-\zeta^{\ast}\right)  \hat{A}_{f}+\left(  1-\zeta\right)
\hat{A}_{f^{\ast}}^{\dagger}}{1-\left\vert \zeta\right\vert ^{2}},\text{
}\label{48}\\
&  \hat{P}=\hbar\frac{\hat{a}-\hat{a}^{\dagger}}{i\sqrt{2}l}=\frac{\hbar
}{i\sqrt{2}l}\frac{\left(  1+\zeta^{\ast}\right)  \hat{A}_{f}-\left(
1+\zeta\right)  \hat{A}_{f^{\ast}}^{\dagger}}{1-\left\vert \zeta\right\vert
^{2}},\nonumber
\end{align}
where $l$ is a length-dimensional parameter which is related to the initial
standard deviation \cite{PER2021}. The new operator $\hat{A}_{f}$ acts on
states $\left\vert \zeta,\xi\right\rangle $ as follows
\begin{equation}
\hat{A}_{f}\left\vert \zeta,\xi\right\rangle =\xi\left\vert \zeta
,\xi\right\rangle . \label{48a}%
\end{equation}
Using the relations (\ref{48}) and (\ref{48a}), we can easily calculate the
mean values of the operators $\hat{x}$ and $\hat{P}$,
\begin{align}
\overline{x}  &  =\overline{x}\left(  t\right)  =\left\langle \xi
,\zeta\left\vert \hat{x}\right\vert \zeta,\xi\right\rangle =\sqrt{2}%
l\frac{\operatorname{Re}\left[  \left(  1-\zeta^{\ast}\right)  \xi\right]
}{1-\left\vert \zeta\right\vert ^{2}},\nonumber\\
\overline{P}  &  =\overline{P}\left(  t\right)  =\left\langle \xi
,\zeta\left\vert \hat{P}\right\vert \zeta,\xi\right\rangle =\frac{\sqrt
{2}\hbar}{l}\frac{\operatorname{Im}\left[  \left(  1+\zeta^{\ast}\right)
\xi\right]  }{1-\left\vert \zeta\right\vert ^{2}}. \label{49}%
\end{align}
From here, there is a correspondence between the squeeze $\zeta$ and
displacement $\xi$ parameters with the mean values of $\overline{x}$ and
$\overline{P}$,
\begin{equation}
\xi=\frac{1+\zeta}{\sqrt{2}l}\overline{x}+\frac{il}{\hbar}\frac{1-\zeta}%
{\sqrt{2}}\overline{P}. \label{49a}%
\end{equation}

Taking the square of the operators $\hat{x}$ and $\hat{P}$, we have
\begin{align}
&  \hat{x}^{2}=\frac{l^{2}}{2}\frac{\left(  1-\zeta^{\ast}\right)  ^{2}\hat
{A}_{f}^{2}+\left(  1-\zeta\right)  ^{2}\hat{A}_{f^{\ast}}^{\dagger
2}+2\left\vert 1-\zeta\right\vert ^{2}\hat{A}_{f^{\ast}}^{\dagger}\hat{A}_{f}%
}{\left(  1-\left\vert \zeta\right\vert ^{2}\right)  ^{2}}+l^{2}%
\frac{\left\vert 1-\zeta\right\vert ^{2}}{1-\left\vert \zeta\right\vert ^{2}%
}\frac{1+\left(  2\varepsilon-1\right)  \hat{R}}{2},\nonumber\\
&  \hat{P}^{2}=-\frac{\hbar^{2}}{2l^{2}}\frac{\left(  1+\zeta^{\ast}\right)
^{2}\hat{A}_{f}^{2}+\left(  1+\zeta\right)  ^{2}\hat{A}_{f^{\ast}}^{\dagger
2}-2\left\vert 1+\zeta\right\vert ^{2}\hat{A}_{f^{\ast}}^{\dagger}\hat{A}_{f}%
}{\left(  1-\left\vert \zeta\right\vert ^{2}\right)  ^{2}}+\frac{\hbar^{2}%
}{l^{2}}\frac{\left\vert 1+\zeta\right\vert ^{2}}{1-\left\vert \zeta
\right\vert ^{2}}\frac{1+\left(  2\varepsilon-1\right)  \hat{R}}{2}.
\label{50}%
\end{align}
So, from these results, we can calculate the mean values of $\overline{x^{2}}$
and $\overline{P^{2}}$, as follow
\begin{align}
\overline{x^{2}}  &  =\overline{x^{2}}\left(  t\right)  =\left\langle
\xi,\zeta\left\vert \hat{x}^{2}\right\vert \zeta,\xi\right\rangle
=\frac{2l^{2}\operatorname{Re}^{2}\left[  \left(  1-\zeta^{\ast}\right)
\xi\right]  }{\left(  1-\left\vert \zeta\right\vert ^{2}\right)  ^{2}}%
+l^{2}\frac{\left\vert 1-\zeta\right\vert ^{2}}{1-\left\vert \zeta\right\vert
^{2}}\frac{1+\left(  2\varepsilon-1\right)  \overline{R}}{2},\nonumber\\
\overline{P^{2}}  &  =\overline{P^{2}}\left(  t\right)  =\left\langle
\xi,\zeta\left\vert \hat{P}^{2}\right\vert \zeta,\xi\right\rangle
=\frac{2\hbar^{2}\operatorname{Im}^{2}\left[  \left(  1+\zeta^{\ast}\right)
\xi\right]  }{l^{2}\left(  1-\left\vert \zeta\right\vert ^{2}\right)  ^{2}%
}+\frac{\hbar^{2}}{l^{2}}\frac{\left\vert 1+\zeta\right\vert ^{2}%
}{1-\left\vert \zeta\right\vert ^{2}}\frac{1+\left(  2\varepsilon-1\right)
\overline{R}}{2}, \label{51}%
\end{align}
where
\begin{equation}
\overline{R}=\left\langle \xi,\zeta\left\vert \hat{R}\right\vert \zeta
,\xi\right\rangle =\frac{I_{\varepsilon-1}\left(  \frac{\left\vert
\xi\right\vert ^{2}}{1-\left\vert \zeta\right\vert ^{2}}\right)
-I_{\varepsilon}\left(  \frac{\left\vert \xi\right\vert ^{2}}{1-\left\vert
\zeta\right\vert ^{2}}\right)  }{I_{\varepsilon-1}\left(  \frac{\left\vert
\xi\right\vert ^{2}}{1-\left\vert \zeta\right\vert ^{2}}\right)
+I_{\varepsilon}\left(  \frac{\left\vert \xi\right\vert ^{2}}{1-\left\vert
\zeta\right\vert ^{2}}\right)  }. \label{52}%
\end{equation}
Note that $\xi=0$ leads to a well-defined even parity of the states
(\ref{30}), once $\overline{R}=1$.

In what follows, we calculate the standard deviation, i.e.,
\begin{align}
\sigma_{x}  &  =\sigma_{x}\left(  t\right)  =\sqrt{\overline{x^{2}}%
-\overline{x}^{2}}=l\left\vert 1-\zeta\right\vert \sqrt{\frac{1+\left(
2\varepsilon-1\right)  \overline{R}}{2\left(  1-\left\vert \zeta\right\vert
^{2}\right)  }},\nonumber\\
\sigma_{P}  &  =\sigma_{P}\left(  t\right)  =\sqrt{\overline{P^{2}}%
-\overline{P}^{2}}=\frac{\hbar}{l}\left\vert 1+\zeta\right\vert \sqrt
{\frac{1+\left(  2\varepsilon-1\right)  \overline{R}}{2\left(  1-\left\vert
\zeta\right\vert ^{2}\right)  }}. \label{53}%
\end{align}
From Eqs. (\ref{53}), it is easy to see that standard deviations in position
and momentum present the squeezing property.

Finally, we aim to find the Heisenberg uncertainty relation in the generalized
states (\ref{30}) by considering the WHA. Therefore, in the hold of previous
results, we can explicitly calculate the product $\sigma_{x}\sigma_{P}$, as
shown below:
\begin{equation}
\sigma_{x}\sigma_{P}=\hbar\frac{\left\vert 1-\zeta\right\vert \left\vert
1+\zeta\right\vert }{1-\left\vert \zeta\right\vert ^{2}}\frac{1+\left(
2\varepsilon-1\right)  \overline{R}}{2}=\hbar\sqrt{1+\frac{4\operatorname{Im}%
^{2}\left(  \zeta\right)  }{\left(  1-\left\vert \zeta\right\vert ^{2}\right)
^{2}}}\frac{1+\left(  2\varepsilon-1\right)  \overline{R}}{2}. \label{54}%
\end{equation}
In particular, if we consider that the squeeze parameter assumes real values,
this implies that $\sigma_{x}\sigma_{P}=\hbar/2\left(  1+\left(
2\varepsilon-1\right)  \overline{R}\right)  $. In turn, for this particular
choice, we should note that the Heisenberg uncertainty assumes the minimum
value predicted by the relation
\begin{equation}
\sigma_{A}\sigma_{B}\geq\frac{1}{2}\left\vert \left\langle \left[  \hat
{A},\hat{B}\right]  \right\rangle \right\vert . \label{uncert}%
\end{equation}
It must be highlighted that in Ref. \citep{SHA1981}, a detailed analysis has
been applied to show how the para-Bose uncertainty relation leads to
uncertainty relation computed from canonical commutation relation.

Taking into account the covariance $\sigma_{xP}$,
\begin{equation}
\sigma_{xP}=\frac{\left\langle \xi,\zeta\left\vert \hat{P}\hat{x}\right\vert
\zeta,\xi\right\rangle +\left\langle \xi,\zeta\left\vert \hat{x}\hat
{P}\right\vert \zeta,\xi\right\rangle }{2}-\overline{x}\left(  t\right)
\overline{P}\left(  t\right)  =-\hbar\operatorname{Im}\left(  \zeta\right)
\frac{1+\left(  2\varepsilon-1\right)  \overline{R}}{1-\left\vert
\zeta\right\vert ^{2}}, \label{55}%
\end{equation}
we can calculate the Schr\"{o}dinger-Robertson uncertainty relation
\citep{ROB1930},
\begin{equation}
\sigma_{x}^{2}\sigma_{P}^{2}-\sigma_{xP}^{2}=\frac{\hbar^{2}}{4}\left[
1+\left(  2\varepsilon-1\right)  \overline{R}\right]  ^{2}, \label{56}%
\end{equation}
which, as can we see, is minimized.

\section{Coordinate representation of the generalized CS\label{V}}

It is well-known that the phase space in the context of quantum mechanics
encounters difficulties due to the uncertainty principle. Considering the
states $\left\vert \zeta,\xi\right\rangle $ in a coordinate representation
allow us to introduce the quasiprobability Wigner distribution, which plays an
analogous role to the classical distributions \citep{HIL1984,WIL2007,OLI2012}.

According to WHA, $\hat{P}$ is a self-adjoint operator on semi-axis $\left(
x\geq0\right)  $, see \citep{MUK1980}, and have the following form:
\begin{equation}
\hat{P}=-i\hbar\partial_{x}+\frac{i\hbar}{2x}\left(  2\varepsilon-1\right)
\hat{R}. \label{31}%
\end{equation}
In this case, the annihilation operator takes the form
\begin{equation}
\hat{a}=\frac{1}{\sqrt{2}}\left(  \frac{\hat{x}}{l}+\frac{il}{\hbar}\hat
{P}\right)  =\frac{l}{\sqrt{2}}\left(  \partial_{x}-\frac{2\varepsilon-1}%
{2x}\hat{R}+\frac{x}{l^{2}}\right)  . \label{32}%
\end{equation}

Applying the annihilation condition
\begin{equation}
\hat{a}\Psi_{0,\varepsilon}\left(  x\right)  =0,\text{ \ }\Psi_{0,\varepsilon
}\left(  x\right)  =\left\langle x|0,\varepsilon\right\rangle , \label{32a}%
\end{equation}
we can obtain a differential equation for the vacuum state, as follows:
\begin{equation}
\left(  \partial_{x}+\frac{x}{l^{2}}-\frac{2\varepsilon-1}{2x}\right)
\Psi_{0,\varepsilon}\left(  x\right)  =0,\text{ \ }\hat{R}\Psi_{0,\varepsilon
}\left(  x\right)  =\Psi_{0,\varepsilon}\left(  x\right)  . \label{33}%
\end{equation}
The general solution reads
\begin{equation}
\Psi_{0,\varepsilon}\left(  x\right)  =Cx^{\varepsilon-\frac{1}{2}}\exp\left(
-\frac{x^{2}}{2l^{2}}\right)  , \label{34}%
\end{equation}
with $C$ being a real constant, which will be determined through the
normalization condition, as seen below,
\begin{equation}
2C^{2}\int_{0}^{\infty}x^{2\varepsilon-1}\exp\left(  -\frac{x^{2}}{l^{2}%
}\right)  dx=1\Rightarrow C=\frac{1}{l^{\varepsilon}\sqrt{\Gamma\left(
\varepsilon\right)  }}. \label{35}%
\end{equation}
It should be noted that the condition (\ref{33}) leads to the following
quantization condition:
\begin{equation}
\varepsilon=2\ell+\frac{1}{2},\text{ \ }\ell=0,1,2,\ldots, \label{37}%
\end{equation}
where $\ell$ is analogous to the angular momentum. Such quantization was
obtained in \citep{LOH2004} to ensure that the eigenfunctions of $\hat{P}$ are
differentiable at the origin.

Thus, the vacuum state with even parity $\Psi_{0,\varepsilon}\left(  x\right)
\equiv\Psi_{0,\ell}^{e}\left(  x\right)  $ takes the form
\begin{equation}
\Psi_{0,\ell}^{e}\left(  x\right)  =\frac{1}{l^{2\ell+\frac{1}{2}}\sqrt
{\Gamma\left(  2\ell+\frac{1}{2}\right)  }}x^{2\ell}\exp\left(  -\frac{x^{2}%
}{2l^{2}}\right)  . \label{38}%
\end{equation}
Taking into account that $\Psi_{\zeta,\xi}^{\ell}\left(  x,t\right)
=\left\langle x|\zeta,\xi\right\rangle $ and replacing the relations
(\ref{10}) in the states (\ref{30}), we obtain
\begin{align}
&  \Psi_{\zeta,\xi}^{\ell}\left(  x,t\right)  =\xi^{2\ell-\frac{1}{2}}%
\frac{\sqrt{\left(  1-\left\vert \zeta\right\vert ^{2}\right)  \Gamma\left(
2\ell+\frac{1}{2}\right)  }\exp\left[  \frac{\zeta^{\ast}\xi^{2}}{2\left(
1-\left\vert \zeta\right\vert ^{2}\right)  }+i\tilde{\vartheta}\right]
}{2^{\frac{4\ell-1}{4}}\sqrt{I_{2\ell-\frac{1}{2}}\left(  \frac{\left\vert
\xi\right\vert ^{2}}{1-\left\vert \zeta\right\vert ^{2}}\right)
+I_{2\ell+\frac{1}{2}}\left(  \frac{\left\vert \xi\right\vert ^{2}%
}{1-\left\vert \zeta\right\vert ^{2}}\right)  }}\nonumber\\
&  \times%
{\displaystyle\sum\limits_{n=0}^{\infty}}
\left(  -\frac{\zeta\hat{a}^{\dagger2}}{2}\right)  ^{n}\left[  \frac
{L_{n}^{2\ell-\frac{1}{2}}\left(  \frac{\xi^{2}}{2\zeta}\right)  }%
{\Gamma\left(  n+2\ell+\frac{1}{2}\right)  }+\frac{L_{n}^{2\ell+\frac{1}{2}%
}\left(  \frac{\xi^{2}}{2\zeta}\right)  }{\Gamma\left(  n+2\ell+\frac{3}%
{2}\right)  }\frac{\xi\hat{a}^{\dagger}}{2}\right]  \Psi_{0,\ell}^{e}\left(
x\right)  . \label{40}%
\end{align}
Using the results below
\begin{align}
&  \left(  \hat{a}^{\dagger}\right)  ^{2n}\Psi_{0,\ell}^{e}\left(  x\right)
=\left(  -1\right)  ^{n}2^{n}n!L_{n}^{2\ell-\frac{1}{2}}\left(  \frac{x^{2}%
}{l^{2}}\right)  \Psi_{0,\ell}^{e}\left(  x\right)  ,\nonumber\\
&  \left(  \hat{a}^{\dagger}\right)  ^{2n+1}\Psi_{0,\ell}^{e}\left(  x\right)
=\frac{\sqrt{2}x}{l}\left(  -1\right)  ^{n}2^{n}n!L_{n}^{2\ell+\frac{1}{2}%
}\left(  \frac{x^{2}}{l^{2}}\right)  \Psi_{0,\ell}^{e}\left(  x\right)  ,
\label{41}%
\end{align}
we can write (\ref{40}), as follows
\begin{align}
&  \Psi_{\zeta,\xi}^{\ell}\left(  x,t\right)  =\left\langle x|\zeta
,\xi\right\rangle =\frac{\sqrt{1-\left\vert \zeta\right\vert ^{2}}}{1-\zeta
}\frac{\sqrt{x}}{l}\frac{I_{2\ell-\frac{1}{2}}\left(  \frac{\sqrt{2}\xi
}{1-\zeta}\frac{x}{l}\right)  +I_{2\ell+\frac{1}{2}}\left(  \frac{\sqrt{2}\xi
}{1-\zeta}\frac{x}{l}\right)  }{\sqrt{I_{2\ell-\frac{1}{2}}\left(
\frac{\left\vert \xi\right\vert ^{2}}{1-\left\vert \zeta\right\vert ^{2}%
}\right)  +I_{2\ell+\frac{1}{2}}\left(  \frac{\left\vert \xi\right\vert ^{2}%
}{1-\left\vert \zeta\right\vert ^{2}}\right)  }}\nonumber\\
&  \times\exp\left[  -\frac{1+\zeta}{1-\zeta}\frac{x^{2}}{2l^{2}}%
-\frac{\left(  1-\zeta^{\ast}\right)  \xi^{2}}{2\left(  1-\zeta\right)
\left(  1-\left\vert \zeta\right\vert ^{2}\right)  }+i\tilde{\vartheta
}\right]  . \label{42}%
\end{align}

In particular, if $\ell=0$ we obtain the following result:
\begin{align}
&  \Psi_{\zeta,\xi}^{0}\left(  x,t\right)  =\frac{\left(  1-\left\vert
\zeta\right\vert ^{2}\right)  ^{1/4}}{\sqrt{\sqrt{\pi}l\left(  1-\zeta\right)
}}\exp\left[  -\frac{1}{2l^{2}}\frac{1+\zeta}{1-\zeta}\left(  x-\frac
{l\sqrt{2}\xi}{1+\zeta}\right)  ^{2}+\frac{\left(  1+\zeta^{\ast}\right)
}{\left(  1+\zeta\right)  \left(  1-\left\vert \zeta\right\vert ^{2}\right)
}\frac{\xi^{2}}{2}-\frac{1}{2}\frac{\left\vert \xi\right\vert ^{2}%
}{1-\left\vert \zeta\right\vert ^{2}}+i\varrho\right]  ,\nonumber\\
&  \varrho=\frac{1}{2}\int\left[  \operatorname{Re}\left(  \alpha\zeta^{\ast
}-\beta\right)  -2\delta\right]  dt. \label{42a}%
\end{align}
Performing the following identifications $\zeta=g/f$, $\xi=z/f=-\varphi/f$ and
$\mu=\left\vert f\right\vert ^{2}\left(  1-\left\vert \zeta\right\vert
^{2}\right)  =1$ leads to the results of the recent publication \citep{PER2021}.

\begin{figure}[ptb]
\begin{centering}
\includegraphics[width=8cm,height=5cm]{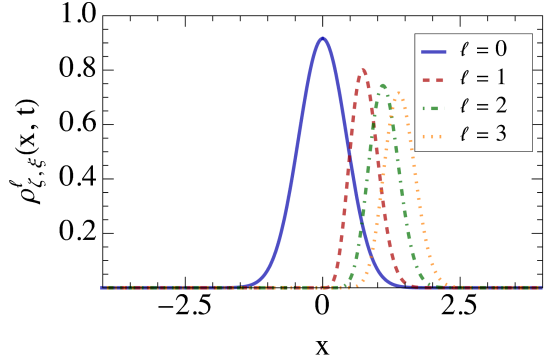}
\par\end{centering}
\caption{Displayed is the probability density by assuming fixed values $l=1$,
$\zeta=0.45$ and $\xi=i$. For $\ell=0$, we have the SCS, while for $\ell>0$,
we obtained the generalized para-Bose CS.}%
\label{fig4}%
\end{figure}

The probability density that corresponds to the generalized CS is given by
\begin{align}
&  \rho_{\zeta,\xi}^{\ell}\left(  x,t\right)  =\left\vert \Psi_{\zeta,\xi
}^{\ell}\left(  x,t\right)  \right\vert ^{2}=\frac{1-\left\vert \zeta
\right\vert ^{2}}{\left\vert 1-\zeta\right\vert ^{2}}\frac{x}{l^{2}}%
\frac{\left\vert I_{2\ell-\frac{1}{2}}\left(  \frac{\sqrt{2}\xi}{1-\zeta}%
\frac{x}{l}\right)  +I_{2\ell+\frac{1}{2}}\left(  \frac{\sqrt{2}\xi}{1-\zeta
}\frac{x}{l}\right)  \right\vert ^{2}}{\left\vert I_{2\ell-\frac{1}{2}}\left(
\frac{\left\vert \xi\right\vert ^{2}}{1-\left\vert \zeta\right\vert ^{2}%
}\right)  +I_{2\ell+\frac{1}{2}}\left(  \frac{\left\vert \xi\right\vert ^{2}%
}{1-\left\vert \zeta\right\vert ^{2}}\right)  \right\vert }\nonumber\\
&  \times\exp\left[  -\frac{1-\left\vert \zeta\right\vert ^{2}}{\left\vert
1-\zeta\right\vert ^{2}}\frac{x^{2}}{l^{2}}-\frac{1}{1-\left\vert
\zeta\right\vert ^{2}}\operatorname{Re}\left(  \frac{1-\zeta^{\ast}}{1-\zeta
}\xi^{2}\right)  \right]  . \label{42b}%
\end{align}
In Fig. \ref{fig4}, we have obtained some plots of the probability density. As
we can see, the probability density of the generalized CS has a shape of a
Gaussian distribution, which moves in space as $\ell$ increases.

Finally, substituting (\ref{48}) into (\ref{3}), the Hamiltonian takes the
form
\begin{align}
\hat{H}  &  =\frac{1}{2}\hbar\left(  \alpha^{\ast}\hat{a}^{2}+\alpha\hat
{a}^{\dagger2}\right)  +\frac{1}{2}\hbar\beta\left(  \hat{a}^{\dagger}\hat
{a}+\hat{a}\hat{a}^{\dagger}\right)  +\hbar\delta\nonumber\\
&  =\frac{\hat{P}^{2}}{2m}+\frac{1}{2}m\omega^{2}\hat{x}^{2}+\frac{1}{2}%
\Omega\left(  \hat{P}\hat{x}+\hat{x}\hat{P}\right)  +\mathcal{E},\nonumber\\
&  =-\frac{\hbar^{2}}{2m}\partial_{x}^{2}-i\hbar\Omega x\partial_{x}%
+\frac{\hbar^{2}}{2mx^{2}}2\ell\left(  2\ell-\hat{R}\right)  +\frac{1}%
{2}m\omega^{2}x^{2}+\mathcal{E}-\frac{i\hbar\Omega}{2}, \label{43}%
\end{align}
where the time-dependent quantities $m$, $\omega$, $\Omega$ and $\mathcal{E}$
reads
\begin{align}
\frac{1}{m}  &  =\frac{l^{2}}{\hbar}\operatorname{Re}\left(  \beta
-\alpha\right)  ,\text{ \ }m\omega^{2}=\frac{\hbar}{l^{2}}\operatorname{Re}%
\left(  \beta+\alpha\right)  ,\text{ \ }\Omega=\operatorname{Im}\left(
\alpha\right)  ,\text{ \ }\mathcal{E}=\hbar\delta,\nonumber\\
\omega^{2}  &  =\beta^{2}-\operatorname{Re}^{2}\left(  \alpha\right)  .
\label{44}%
\end{align}

From Hamiltonian (\ref{43}), we can describe the physical systems such as
harmonic oscillators, which are described by confining potential $V_{HO}\sim
x^{2}$. We also can establish an analogy of the potential $V_{C}=\frac
{\hbar^{2}}{2mx^{2}}2\ell\left(  2\ell-\hat{R}\right)  $ with the centrifugal
potential. On the other hand, one may identify the potential $V_{C}$ with the
conformal sector of quantum conformal mechanics, except by the $\hat{R}$
operator \cite{AND2016}. Since the reflection operator acting on even parity
states leads to $\hat{R}\psi_{e}=\psi_{e}$, it must be noticed that the
potential $V_{C}$ has a negative correction term owing to the reflection
operator. Furthermore, the repulsiveness of the $V_{C}$ potential is weakened
by the correction arising from the reflection operator.

Finally, we can relate the action of the reflection operator on a particular
state to the reduction of the dimensionality of space. Let us analyze the
centrifugal term of the $d$-dimensional Laplacian, which can be written in the
form \cite{LOH2004}:
\begin{equation}
\Delta^{\left(  d\right)  }=\partial_{r}^{2}+\frac{d-1}{r}\partial_{r}%
+\frac{L\left(  L+d-2\right)  }{r^{2}}. \label{laplacian}%
\end{equation}

At this point, we can establish a direct relationship between the action of
the reflection operator on even parity states and the reduction of the space
from three to one dimension $\left(  d=1\right)  $, from the perspective of
the centrifugal term by considering that $L=2\ell$.

\section{Time-independent para-Bose Oscillator\label{VI}}

Since the results obtained in this work can be applied to both time-dependent
and time-independent physical systems, for the sake of simplicity, we apply
this approach to the study of the time-independent para-Bose oscillator.
First, we obtain the Hamiltonian of the time-independent para-Bose oscillator
applying the following conditions $\alpha=\delta=\Omega=0$ and $\beta
=\beta_{0}$ on the Eq. (\ref{3}), which can be written in the form:
\begin{align}
&  \hat{H}=\frac{1}{2}\hbar\beta_{0}\left(  \hat{a}^{\dagger}\hat{a}+\hat
{a}\hat{a}^{\dagger}\right)  =\frac{\hbar^{2}}{2m_{0}}\left[  -\partial
_{x}^{2}+\frac{2\ell\left(  2\ell-\hat{R}\right)  }{x^{2}}+\frac{m_{0}%
^{2}\omega_{0}^{2}}{\hbar^{2}}x^{2}\right]  ,\nonumber\\
&  \omega_{0}=\beta_{0},\text{ \ }m_{0}=\frac{\hbar}{l^{2}\beta_{0}},
\label{A1}%
\end{align}
where the subindex labels the initial time, and from (\ref{37}) we can write
the para-Bose number states $\left\vert n,\varepsilon\right\rangle
\Longrightarrow\left\vert n,\ell\right\rangle $, such that
\begin{equation}
\hat{H}\left\vert n,\ell\right\rangle =\hbar\omega_{0}\left(  n+2\ell+\frac
{1}{2}\right)  \left\vert n,\ell\right\rangle . \label{A2}%
\end{equation}
Notice that, taking $\ell=0$ lead to the standard harmonic oscillator.

In this case, the equation system (\ref{7}), take the form
\begin{equation}
\dot{f}=i\omega_{0}f,\text{ \ }\dot{g}=-i\omega_{0}g,\text{ \ }\dot{\varphi
}=0, \label{A3}%
\end{equation}
whose solution is given by
\begin{equation}
f=f_{0}e^{i\omega_{0}t},\text{ \ }g=g_{0}e^{-i\omega_{0}t},\text{ \ }%
\varphi\equiv0. \label{A4}%
\end{equation}
From here, we can write the squeeze and displacement parameters in the form
\begin{align}
&  \zeta=\frac{g}{f}=\zeta_{0}e^{-2i\omega_{0}t},\text{ \ }\xi=\frac{z}{f}%
=\xi_{0}e^{-i\omega_{0}t},\nonumber\\
&  \zeta_{0}=\left\vert \zeta_{0}\right\vert e^{i\theta_{\zeta}}=\frac{g_{0}%
}{f_{0}},\text{ \ }\xi_{0}=\left\vert \xi_{0}\right\vert e^{i\theta_{\xi}%
}=\frac{z}{f_{0}}. \label{A5}%
\end{align}
In the following, we can rewrite the operators (\ref{48}):
\begin{align}
&  \hat{a}=\frac{\hat{A}_{f}-\zeta_{0}e^{-2i\omega_{0}t}\hat{A}_{f^{\ast}%
}^{\dagger}}{1-\left\vert \zeta_{0}\right\vert ^{2}},\text{ \ }\hat{A}%
_{f}\equiv\frac{1}{f_{0}e^{i\omega_{0}t}}\hat{A},\nonumber\\
&  \hat{x}=\frac{l}{\sqrt{2}}\frac{\left(  1-\zeta_{0}^{\ast}e^{2i\omega_{0}%
t}\right)  \hat{A}_{f}+\left(  1-\zeta_{0}e^{-2i\omega_{0}t}\right)  \hat
{A}_{f^{\ast}}^{\dagger}}{1-\left\vert \zeta_{0}\right\vert ^{2}},\text{
}\nonumber\\
&  \hat{P}=\frac{\hbar}{i\sqrt{2}l}\frac{\left(  1+\zeta_{0}^{\ast}%
e^{2i\omega_{0}t}\right)  \hat{A}_{f}-\left(  1+\zeta_{0}e^{-2i\omega_{0}%
t}\right)  \hat{A}_{f^{\ast}}^{\dagger}}{1-\left\vert \zeta_{0}\right\vert
^{2}}, \label{A6}%
\end{align}
and the generalized CS (\ref{30}), as follows
\begin{align}
&  \left\vert \zeta,\xi\right\rangle =\left(  \frac{\left\vert \xi
_{0}\right\vert e^{i\theta_{\xi}}}{\sqrt{2}}\right)  ^{2\ell-\frac{1}{2}}%
\sqrt{\frac{1-\left\vert \zeta_{0}\right\vert ^{2}}{I_{2\ell-\frac{1}{2}%
}\left(  \frac{\left\vert \xi_{0}\right\vert ^{2}}{1-\left\vert \zeta
_{0}\right\vert ^{2}}\right)  +I_{2\ell+\frac{1}{2}}\left(  \frac{\left\vert
\xi_{0}\right\vert ^{2}}{1-\left\vert \zeta_{0}\right\vert ^{2}}\right)  }%
}\exp\left[  \frac{\left\vert \zeta_{0}\right\vert \left\vert \xi
_{0}\right\vert ^{2}e^{2i\theta_{\xi}-i\theta_{\zeta}}}{2\left(  1-\left\vert
\zeta_{0}\right\vert ^{2}\right)  }\right]  \times\nonumber\\
&
{\displaystyle\sum\limits_{n=0}^{\infty}}
\left(  -\zeta_{0}\right)  ^{n}\sqrt{n!}e^{-2i\omega_{0}\left(  n+\ell\right)
t}\left[  \frac{L_{n}^{2\ell-\frac{1}{2}}\left(  \frac{\left\vert \xi
_{0}\right\vert ^{2}e^{2i\theta_{\xi}}}{2\left\vert \zeta_{0}\right\vert
e^{i\theta_{\zeta}}}\right)  e^{-\frac{i\omega_{0}t}{2}}}{\sqrt{\Gamma\left(
n+2\ell+\frac{1}{2}\right)  }}\left\vert 2n,\ell\right\rangle +\frac
{\left\vert \xi_{0}\right\vert e^{i\theta_{\xi}}L_{n}^{2\ell+\frac{1}{2}%
}\left(  \frac{\left\vert \xi_{0}\right\vert ^{2}e^{2i\theta_{\xi}}%
}{2\left\vert \zeta_{0}\right\vert e^{i\theta_{\zeta}}}\right)  e^{-\frac
{3i\omega_{0}}{2}t}}{\sqrt{2\Gamma\left(  n+2\ell+\frac{3}{2}\right)  }%
}\left\vert 2n+1,\ell\right\rangle \right]  . \label{A7}%
\end{align}

From Eq. (\ref{49}), we get $\overline{x}$ and $\overline{P}$ in the form
\begin{align}
&  \overline{x}=\overline{x}_{0}\cos\left(  \omega_{0}t\right)  +\frac
{\overline{P}_{0}}{m_{0}\omega_{0}}\sin\left(  \omega_{0}t\right)  ,\text{
\ }\overline{P}=\overline{P}_{0}\cos\left(  \omega_{0}t\right)  -m_{0}%
\omega_{0}\overline{x}_{0}\sin\left(  \omega_{0}t\right)  ,\nonumber\\
&  \overline{x}_{0}=\sqrt{2}l\left\vert \xi_{0}\right\vert \frac{\cos\left(
\theta_{\xi}\right)  -\left\vert \zeta_{0}\right\vert \cos\left(  \theta_{\xi
}-\theta_{\zeta}\right)  }{1-\left\vert \zeta_{0}\right\vert ^{2}},\text{
\ }\overline{P}_{0}=\sqrt{2}lm_{0}\omega_{0}\left\vert \xi_{0}\right\vert
\frac{\sin\left(  \theta_{\xi}\right)  +\left\vert \zeta_{0}\right\vert
\sin\left(  \theta_{\xi}-\theta_{\zeta}\right)  }{1-\left\vert \zeta
_{0}\right\vert ^{2}}. \label{A8}%
\end{align}
Notice that taking $\xi_{0}=0$, the mean values of $\overline{x}$ and
$\overline{P}$ are equal to zero, which corresponds to the mean values
evaluated in the generalized SVS.

The uncertainty relation (\ref{54}) and (\ref{56}) becomes
\begin{align}
&  \sigma_{x}\sigma_{P}=\hbar\sqrt{1+\frac{4\left\vert \zeta_{0}\right\vert
^{2}\sin^{2}\left(  \theta_{\zeta}-2\omega_{0}t\right)  }{\left(  1-\left\vert
\zeta_{0}\right\vert ^{2}\right)  ^{2}}}\frac{1+4\ell\overline{R}}%
{2},\nonumber\\
&  \sigma_{x}^{2}\sigma_{P}^{2}-\sigma_{xP}^{2}=\frac{\hbar^{2}}{4}\left(
1+4\ell\overline{R}\right)  ^{2}, \label{A9}%
\end{align}
with the mean value of $\hat{R}$ is given by the Eq. (\ref{52}),
\begin{equation}
\overline{R}=\frac{I_{2\ell-\frac{1}{2}}\left(  \frac{\left\vert \xi
_{0}\right\vert ^{2}}{1-\left\vert \zeta_{0}\right\vert ^{2}}\right)
-I_{2\ell+\frac{1}{2}}\left(  \frac{\left\vert \xi_{0}\right\vert ^{2}%
}{1-\left\vert \zeta_{0}\right\vert ^{2}}\right)  }{I_{2\ell-\frac{1}{2}%
}\left(  \frac{\left\vert \xi_{0}\right\vert ^{2}}{1-\left\vert \zeta
_{0}\right\vert ^{2}}\right)  +I_{2\ell+\frac{1}{2}}\left(  \frac{\left\vert
\xi_{0}\right\vert ^{2}}{1-\left\vert \zeta_{0}\right\vert ^{2}}\right)  }.
\label{A10}%
\end{equation}
One can see that the uncertainty relation has an oscillatory behavior,
reaching minimum values at specific points given by
\begin{equation}
\sin\left(  \theta_{\zeta}-2\omega_{0}t_{k}\right)  =0,\text{ \ }t_{k}%
=\frac{\theta_{\zeta}-k\pi}{2\omega_{0}}, \label{A11}%
\end{equation}
where $k=0,\pm1,\pm2,\ldots$. Here $t_{k}$ corresponds to the values on the
time for which the uncertainty relation is minimal. The condition
$\theta_{\zeta}=0\Longrightarrow\zeta_{0}=\zeta_{0}^{\ast}$ leads to minimum
uncertainty at $t=0$. From Eqs. (\ref{53}), (\ref{37}) and (\ref{A9}), we can
write the $l$-parameter in term of the quantities $\sigma_{x}\left(  0\right)
=\sigma_{x_{0}}$, $\zeta_{0}$, $\xi_{0}$ and $\ell$, as follows
\begin{equation}
l=\sqrt{\frac{1+\zeta_{0}}{1-\zeta_{0}}\frac{2}{1+4\ell\overline{R}}}%
\sigma_{x_{0}},\text{ \ }\sigma_{x_{0}}=\hbar\frac{1+4\ell\overline{R}%
}{2\sigma_{P_{0}}}. \label{A11a}%
\end{equation}
In what the following, let us consider $\zeta_{0}$ as being a real parameter.

\begin{figure}[ptb]
\subfloat[]{\includegraphics[width=8cm,height=5cm]{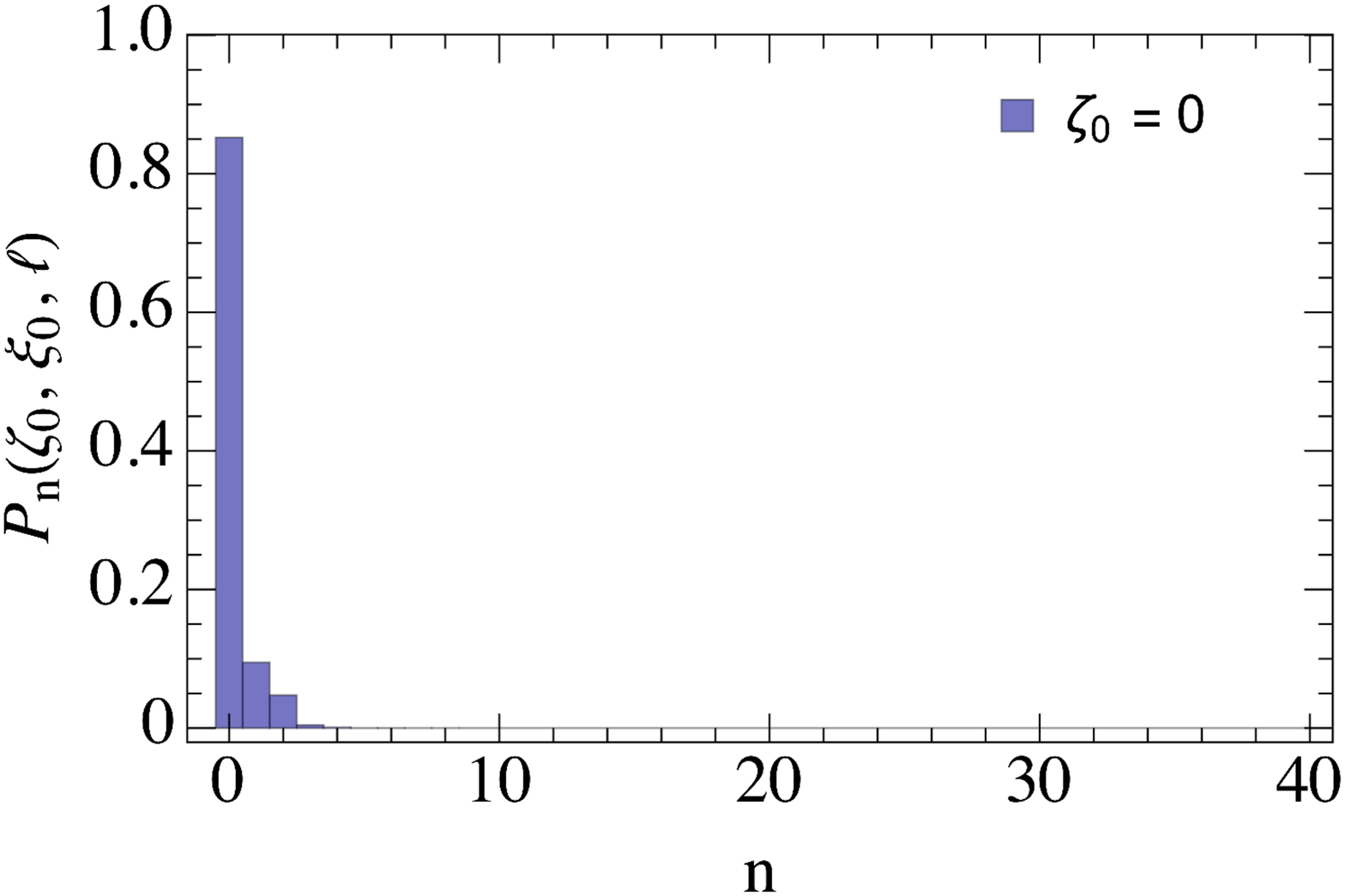}
}\hfill{}\subfloat[]{\includegraphics[width=8cm,height=5cm]{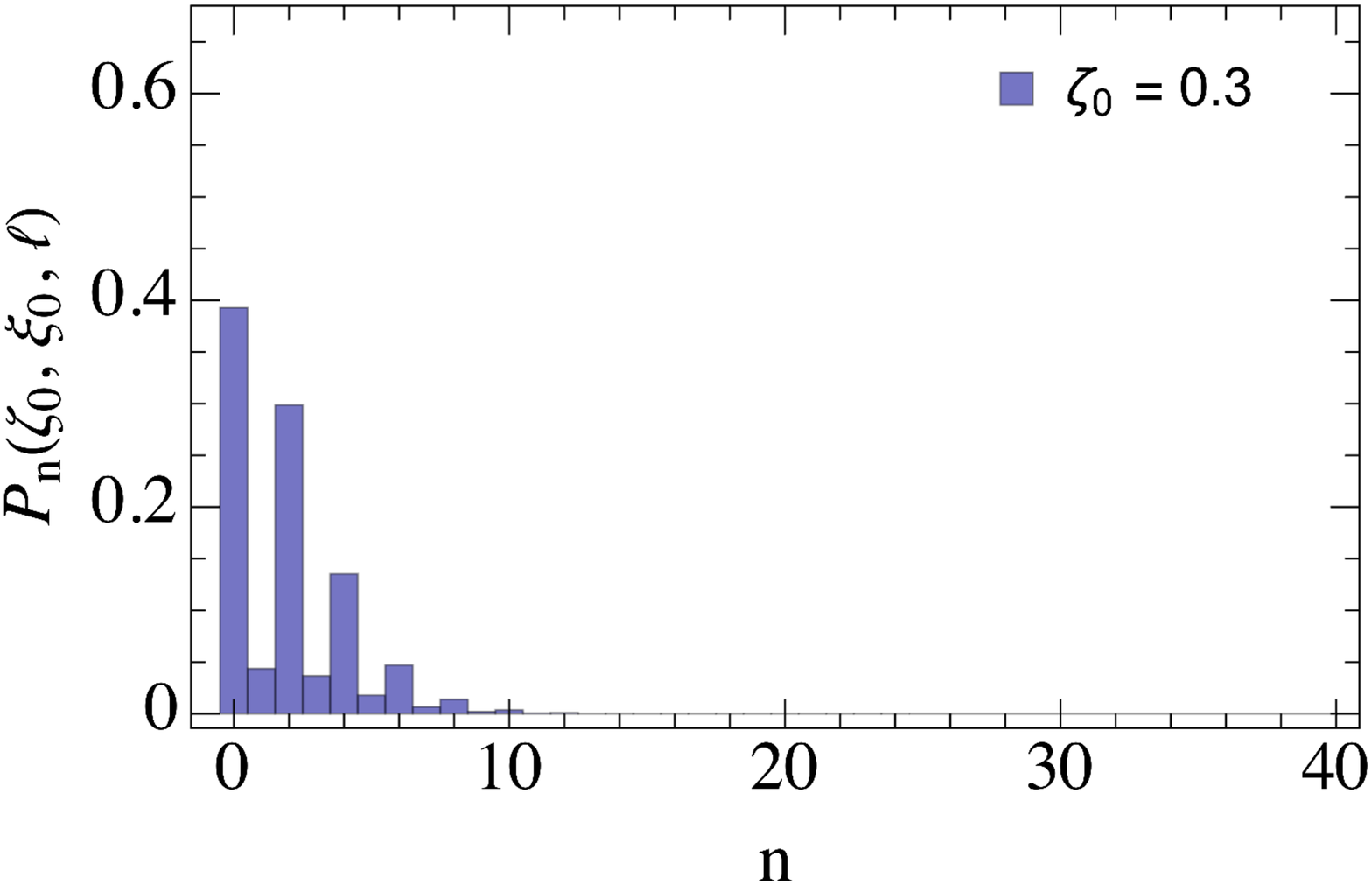}
}
\par
\vfill{}
\par
\subfloat[]{\includegraphics[width=8cm,height=5cm]{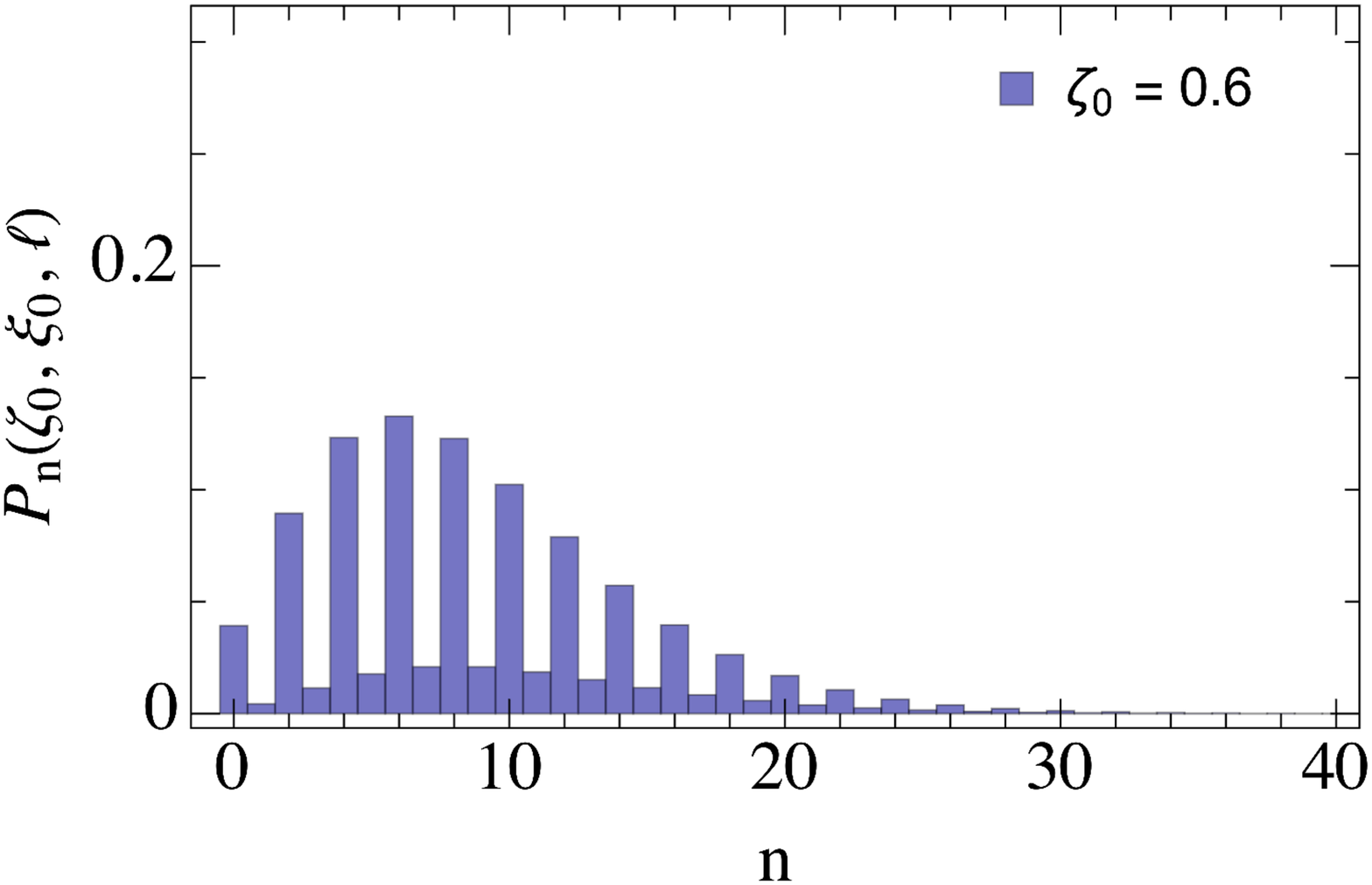}
}\hfill{}\subfloat[]{\includegraphics[width=8cm,height=5cm]{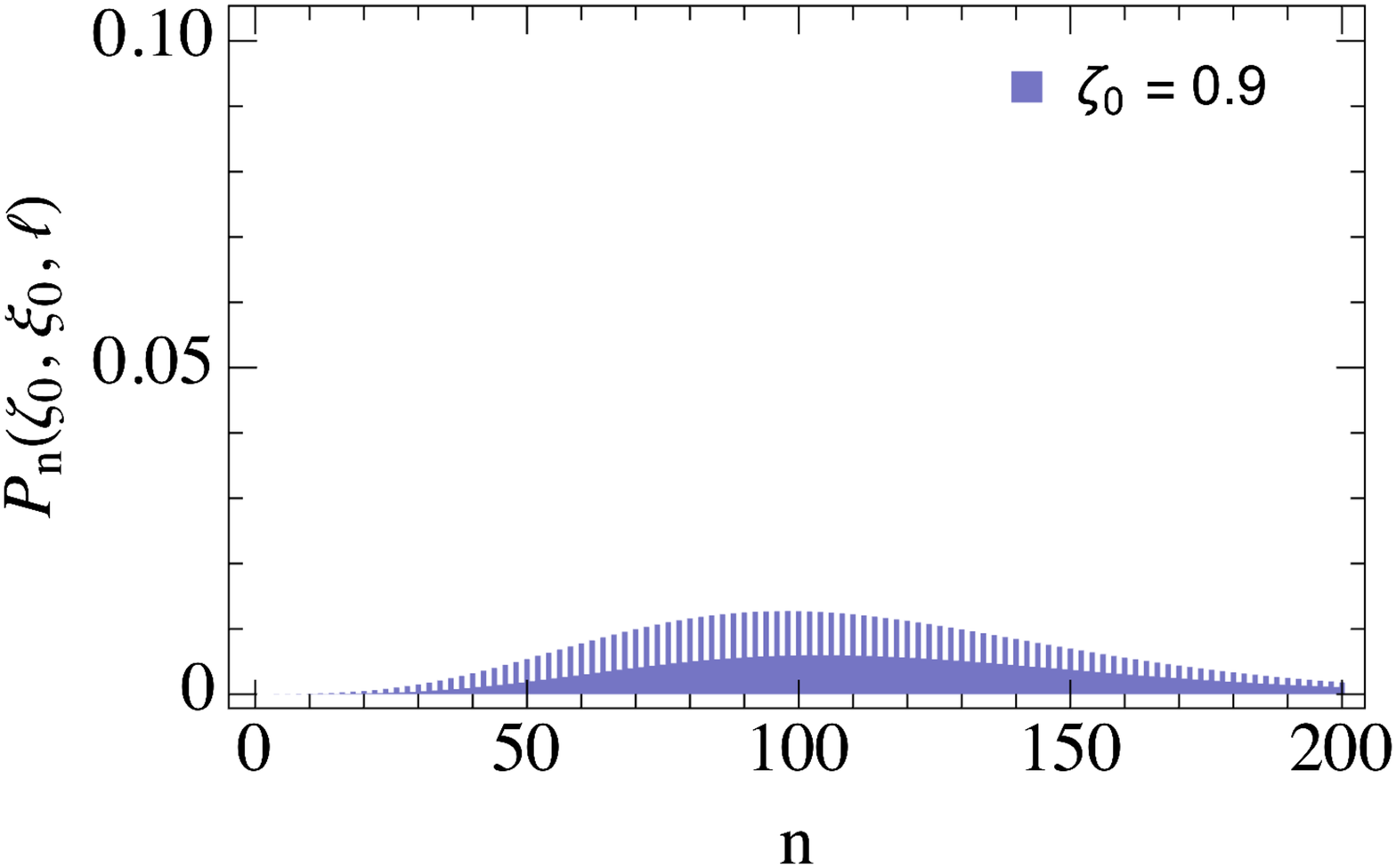}
}\caption{The probability transition is showed by considering fixed values
$\ell=2$, $\left\vert \xi_{0}\right\vert =1$, and $\theta_{\xi}=\pi/2$. These
conditions imply that $\overline{x}_{0}=0$ and $\overline{P}_{0}=\sqrt
{2}lm_{0}\omega_{0}\left\vert \xi_{0}\right\vert \left(  1+\zeta_{0}\right)
/\left(  1-\zeta_{0}^{2}\right)  $. In Figure (a) we recover the para-Bose CS.
Meanwhile, in Figures (b), (c) and (d) we have the generalized para-Bose CS.}%
\label{fig5}%
\end{figure}

It must be highlighted that the para-Bose number states are eigenstates of the
Hamiltonian (\ref{A1}), and therefore the standard deviation for this operator
is null when evaluated on this basis. Since $\hat{H}$ is time-independent, we
have that $\left\vert n,\ell\right\rangle $ are eigenstates of $\hat{H}$ with
well-defined energy. Furthermore, it is interesting to verify the probability
transition from the states with well-defined energy to generalized CS
(\ref{A7}). Therefore, it follows from (\ref{30a}), (\ref{37}) and (\ref{A5})
that
\begin{align}
&  P_{n}\left(  \zeta_{0},\xi_{0},\ell\right)  =\left(  \frac{\left\vert
\xi_{0}\right\vert ^{2}}{2}\right)  ^{2\ell-\frac{1}{2}}\frac{\left(
1-\zeta_{0}^{2}\right)  \exp\left[  \frac{\zeta_{0}\left\vert \xi
_{0}\right\vert ^{2}}{1-\zeta_{0}^{2}}\cos\left(  2\theta_{\xi}\right)
\right]  }{I_{2\ell-\frac{1}{2}}\left(  \frac{\left\vert \xi_{0}\right\vert
^{2}}{1-\zeta_{0}^{2}}\right)  +I_{2\ell+\frac{1}{2}}\left(  \frac{\left\vert
\xi_{0}\right\vert ^{2}}{1-\zeta_{0}^{2}}\right)  }n!\zeta_{0}^{2n}\nonumber\\
&  \times\left[  \frac{\left\vert L_{n}^{2\ell-\frac{1}{2}}\left(
\frac{\left\vert \xi_{0}\right\vert ^{2}e^{2i\theta_{\xi}}}{2\zeta_{0}%
}\right)  \right\vert ^{2}}{\Gamma\left(  n+2\ell+\frac{1}{2}\right)  }%
+\frac{\left\vert \xi_{0}\right\vert ^{2}\left\vert L_{n}^{2\ell+\frac{1}{2}%
}\left(  \frac{\left\vert \xi_{0}\right\vert ^{2}e^{2i\theta_{\xi}}}%
{2\zeta_{0}}\right)  \right\vert ^{2}}{2\Gamma\left(  n+2\ell+\frac{3}%
{2}\right)  }\right]  .\label{A12}%
\end{align}
Notice that the probability transition is time-independent. As we saw in the
figures Fig. \ref{fig1} and Fig. \ref{fig3}, the probability transition has
its shape significantly altered by the displacement and Wigner parameters. In
Fig. \ref{fig5}, we will make an analysis considering some values for the
squeeze parameter $\zeta_{0}$, keeping the other parameters fixed; namely,
$\xi_{0}$ and $\ell$.

From Eqs. (\ref{A8}), we have that the $\zeta_{0}$,$\xi_{0}$-parameters are
directly related to the initial conditions that lead to the temporal evolution
of the mean values in the position and momentum. As we saw, the standard
deviation and the uncertainty relations are expressed in terms of special
functions, whose analysis is not straightforward. Therefore, it is interesting
to investigate the asymptotic form of Bessel functions of the first kind
present in uncertainty relations as follows:

a) Asymptotic form for small arguments $Z$ and fixed $\kappa$ (see Eq. 9.6.7,
page 375 in Ref. \cite{Abr1965}).
\begin{equation}
I_{\kappa}\left(  Z\right)  \sim\frac{1}{\Gamma\left(  \kappa+1\right)
}\left(  \frac{Z}{2}\right)  ^{\kappa}. \label{A13}%
\end{equation}
This limit implies that $\left\vert \xi_{0}\right\vert \rightarrow0$ and
$\zeta_{0}<1$. In its turn, the mean value of the reflection operator
$\overline{R}$ can be rewritten:
\begin{equation}
\overline{R}\sim\frac{\left(  4\ell+1\right)  \left(  1-\zeta_{0}^{2}\right)
-\left\vert \xi_{0}\right\vert ^{2}}{\left(  4\ell+1\right)  \left(
1-\zeta_{0}^{2}\right)  +\left\vert \xi_{0}\right\vert ^{2}}, \label{A14}%
\end{equation}
and the uncertainty relations (\ref{A9}) become
\begin{align}
&  \sigma_{x}\sigma_{P}\sim\frac{\hbar}{2}\sqrt{1+\frac{4\zeta_{0}^{2}\sin
^{2}\left(  2\omega_{0}t\right)  }{\left(  1-\zeta_{0}^{2}\right)  ^{2}}}%
\frac{\left(  4\ell+1\right)  ^{2}\left(  1-\zeta_{0}^{2}\right)  -\left(
4\ell-1\right)  \left\vert \xi_{0}\right\vert ^{2}}{\left(  4\ell+1\right)
\left(  1-\zeta_{0}^{2}\right)  +\left\vert \xi_{0}\right\vert ^{2}%
},\nonumber\\
&  \sigma_{x}^{2}\sigma_{P}^{2}-\sigma_{xP}^{2}\sim\frac{\hbar^{2}}{4}\left[
\frac{\left(  4\ell+1\right)  ^{2}\left(  1-\zeta_{0}^{2}\right)  -\left(
4\ell-1\right)  \left\vert \xi_{0}\right\vert ^{2}}{\left(  4\ell+1\right)
\left(  1-\zeta_{0}^{2}\right)  +\left\vert \xi_{0}\right\vert ^{2}}\right]
^{2}. \label{A15}%
\end{align}
From (\ref{A14}), it is easy to see that the parity of the states is even when
we take $\left\vert \xi_{0}\right\vert =0$. On the other hand, the value of
the uncertainties, Eq. (\ref{A15}), increases as $\ell$ increases, while for
$\zeta_{0}=\ell=0$ is minimized.

b) Asymptotic form for large arguments $Z$ and fixed $\kappa$ (see Eq. 9.7.1,
page 377 in Ref. \cite{Abr1965}),
\begin{equation}
I_{\kappa}\left(  Z\right)  \sim\frac{e^{Z}}{\sqrt{2\pi Z}}\left[  1-\frac
{1}{2Z}\left(  \kappa^{2}-\frac{1}{4}\right)  \right]  . \label{A16}%
\end{equation}
This limit can be obtained in two different ways, first it is $\left\vert
\xi_{0}\right\vert \rightarrow\infty$ with $\zeta_{0}<1$ and second $\zeta
_{0}\rightarrow1$ with $\left\vert \xi_{0}\right\vert <\infty$. In this second
case, we have a high degree of squeeze in $\sigma_{x}$, which becomes smaller
compared to $\sigma_{P}$, as $\zeta_{0}$ approaches $1$. From Eq. (\ref{A16}),
we can write the mean value of $\overline{R}$ as follows
\begin{equation}
\overline{R}\sim\frac{\ell\left(  1-\zeta_{0}^{2}\right)  }{\left\vert \xi
_{0}\right\vert ^{2}-2\ell^{2}\left(  1-\zeta_{0}^{2}\right)  }. \label{A17}%
\end{equation}
Note that the range of values $\left\vert \xi_{0}\right\vert ^{2}<2\ell
^{2}\left(  1-\zeta_{0}^{2}\right)  $ takes $\overline{R}<0$, indicating that
only odd parity states stay at this limit. The uncertainty relations can be
rewritten in the form
\begin{align}
&  \sigma_{x}\sigma_{P}\sim\frac{\hbar}{2}\sqrt{1+\frac{4\zeta_{0}^{2}\sin
^{2}\left(  2\omega_{0}t\right)  }{\left(  1-\zeta_{0}^{2}\right)  ^{2}}}%
\frac{\left\vert \xi_{0}\right\vert ^{2}+2\ell^{2}\left(  1-\zeta_{0}%
^{2}\right)  }{\left\vert \xi_{0}\right\vert ^{2}-2\ell^{2}\left(  1-\zeta
_{0}^{2}\right)  },\nonumber\\
&  \sigma_{x}^{2}\sigma_{P}^{2}-\sigma_{xP}^{2}\sim\frac{\hbar^{2}}{4}\left[
\frac{\left\vert \xi_{0}\right\vert ^{2}+2\ell^{2}\left(  1-\zeta_{0}%
^{2}\right)  }{\left\vert \xi_{0}\right\vert ^{2}-2\ell^{2}\left(  1-\zeta
_{0}^{2}\right)  }\right]  ^{2}. \label{A18}%
\end{align}
It is easy to see that $\ell=0$ lead the uncertainty relations to the value
obtained when evaluated in terms of the canonical commutation relation
$\left(  \left[  \hat{x},\hat{p}\right]  =i\hbar\text{, }\hat{p}%
=-i\hbar\partial_{x}\right)  $. Besides, it can be seen that the squeeze
parameter $\zeta_{0}$ ensures that the Heisenberg uncertainty relation
oscillates over time.

This example illustrates how this general procedure can be applied to a range
of problems by simply determining the parameters of the Hamiltonian, then
finding the functions $f$ and $g$. From these functions, we obtain the squeeze
and displacement parameters that, in its turn, modify the uncertainty relation
and probability transition.

\section{Concluding remarks\label{VII}}

In this article, we study the integrals of motion method in a para-Bose
formulation. This approach generalizes the usual canonical commutation
relation. In turn, we obtain a generalization of the usual SVS, which admits a
completeness relation in terms of the Wigner parameter. This relation depends
on a range of values for the Wigner parameter, which does not include that
corresponding to canonical algebra. We also obtain a generalization of the CS
in terms of the even and odd time-independent para-Bose number states. These
states are thoroughly determined in terms of the time-dependent squeeze and
displacement parameters, as well as the Wigner parameter. In the study of the
probability transition, we saw that the displacement parameter has an
additional role, which is a kind of transition parameter by allowing access to
the odd states of the system. Meanwhile, the Wigner parameter has the role of
controlling the \textquotedblleft dispersion," and attenuating the access to
the odd states. We show that the minimization of the Heisenberg uncertainty
relation is easily obtained by taking the real value of the squeeze parameter
and that the squeezing properties can be seen from the standard deviation of
the position and momentum. Taking the coordinate representation of the
generalized CS, we found a quantization condition on the Wigner parameter,
analogous to the quantization of the angular momentum, which arises by
imposing that the parity of the vacuum state is even. This quantization
condition also ensures that the eigenstates of the momentum operator are
differentiable at the origin. Finally, the para-Bose oscillator has been
discussed in this framework.

\begin{acknowledgments}
We would like to thank CNPq, CAPES and CNPq/PRONEX/FAPESQ-PB (Grant no.
165/2018), for partial financial support. ASL and FAB acknowledge support from
CNPq (Grant nos. 150601/2021-2, 312104/2018-9). ASP thanks the support of the
Instituto Federal do Par\'{a}.
\end{acknowledgments}

\textbf{Data Availability Statement}

Data sharing not applicable to this article as no datasets were generated or
analysed during the current study.

\end{document}